\documentclass[10pt,journal, draftcls,onecolumn]{IEEEtran}
\usepackage{epsfig}
\usepackage{graphicx}
\usepackage{amsmath}
\usepackage{amssymb}
\def\mX{\mathcal{X}}
\def\mY{\mathcal{Y}}
\def\mE{\mathcal{E}}
\def\mD{\mathcal{D}}
\def\mC{\mathcal{C}}
\def\mB{\mathcal{B}}
\def\mU{\mathcal{U}}
\def\mV{\mathcal{V}}
\def\Beta{\beta}
\def\wh{\widehat}
\def\wt{\widetilde}
\def\eps{\epsilon}
\def\b{\Big}
\def\B{\bigg}
\def\h{\Bigg}
\def\sr{\stackrel}
\def\brl{\{}
\def\brr{\}}
\def\be{\begin{equation}}
\def\ee{\end{equation}}
\def\ba{\begin{eqnarray}}
\def\ea{\end{eqnarray}}
\newtheorem{example}{Example}[section]
\newtheorem{theorem}{\bf Theorem}
\newtheorem{definition}{\bf Definition}

\begin{document}
\title{Sequential decoding for lossless streaming source coding with side information}%
\author{Hari Palaiyanur, \IEEEmembership{Student Member}, Anant Sahai, \IEEEmembership{Member}
\thanks{This work was supported by the Vodafone US Foundation.}
\thanks{H. Palaiyanur and A. Sahai are with the Electrical Engineering Department at the University of California,
Berkeley.}}
\maketitle

\begin{abstract}
    The problem of lossless fixed-rate streaming coding of discrete memoryless sources with side
information at the decoder is studied. A random time-varying tree-code is used to sequentially
bin strings and a Stack Algorithm with a variable bias uses the side information to give a
delay-universal coding system for lossless source coding with side information. The scheme is
shown to give exponentially decaying probability of error with delay, with exponent equal to
Gallager's random coding exponent for sources with side information. The mean of the random
variable of computation for the stack decoder is bounded, and conditions on the bias are given
to guarantee a finite $\rho^{th}$ moment for $0 \leq \rho \leq 1$.
    Further, the problem is also studied in the case where there is a discrete memoryless
channel between encoder and decoder. The same scheme is slightly modified to give a
joint-source channel encoder and Stack Algorithm-based sequential decoder using side
information. Again, by a suitable choice of bias, the probability of error decays exponentially
with delay and the random variable of computation has a finite mean. Simulation results for
several examples are given.
\end{abstract}


\section{Introduction}

\begin{figure}[hbp]
\begin{center}
\setlength{\unitlength}{3500sp}%
\begingroup\makeatletter\ifx\SetFigFont\undefined%
\gdef\SetFigFont#1#2#3#4#5{%
  \reset@font\fontsize{#1}{#2pt}%
  \fontfamily{#3}\fontseries{#4}\fontshape{#5}%
  \selectfont}%
\fi\endgroup%
\begin{picture}(4842,2039)(946,-2023)
\thicklines \put(1891,-1287){\framebox(945,709){}} \put(2836,-933){\vector( 1, 0){1181}}
\put(1418,-933){\vector( 1, 0){473}} \put(4017,-1287){\framebox(945,709){}}
\put(4962,-933){\vector( 1, 0){708}} \put(4489,{-1796}){\vector( 0, 1){509}}
\put(3350,-1050){\line(1,1){200}}
\put(2250,-1051){\makebox(0,0)[lb]{\smash{{\SetFigFont{14}{16.8}{\rmdefault}{\mddefault}{\updefault}$\mathcal{E}$}}}}
\put(1450,-815){\makebox(0,0)[lb]{\smash{{\SetFigFont{14}{16.8}{\rmdefault}{\mddefault}{\updefault}$U$}}}}
\put(3190,-815){\makebox(0,0)[lb]{\smash{{\SetFigFont{14}{16.8}{\rmdefault}{\mddefault}{\updefault}$R$}}}}
\put(4371,-1051){\makebox(0,0)[lb]{\smash{{\SetFigFont{14}{16.8}{\rmdefault}{\mddefault}{\updefault}$\mathcal{D}$}}}}
\put(4400,-2050){\makebox(0,0)[lb]{\smash{{\SetFigFont{14}{16.8}{\rmdefault}{\mddefault}{\updefault}$V$}}}}
\put(5300,-815){\makebox(0,0)[lb]{\smash{{\SetFigFont{14}{16.8}{\rmdefault}{\mddefault}{\updefault}$\hat{U}$}}}}
\end{picture}
\caption{Source coding with side information at rate $R$ bits per time unit.}
\label{fig:sideinformation}
\end{center}
\end{figure}
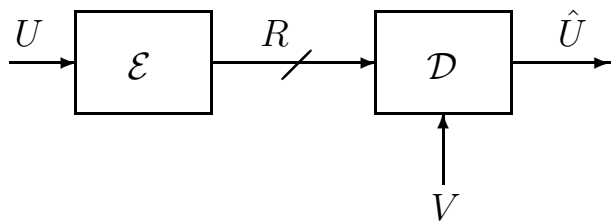

    In this paper, we consider the problem of lossless source coding with side
information shown in Figure \ref{fig:sideinformation}. The seminal paper of Slepian and Wolf
\cite{SlepianWolf} was the first to give the achievable rate region for this problem, when the
source consists of a pair of dependent random variables that are independent and identically
distributed (IID) over time. A sequence of IID symbols is encoded and its compressed
representation is given noiselessly to a decoder. The decoder also has access to side
information that is correlated in a known way with the source. The side information generally
permits the source to be compressed to a rate below its entropy and still recovered losslessly.
If the source is $U$ and the side information $V$, then \cite{SlepianWolf} showed that the
conditional entropy, $H(U|V)$, is a sufficient rate to recover the $U$ with arbitrarily low
probability of error.

    The currently known, robust methods of compression used in point-to-point lossless source coding generally
employ variable length codes. Solutions such as Lempel-Ziv coding (\cite{CoverBook},
\cite{LempelZiv77}, \cite{LempelZiv78}) and context-tree weighting \cite{WillemsContextTree}
are also capable of efficiently compressing many sources with memory. Recently, these
algorithms have been adapted to the `compression with side information' problem when the side
information is available to both the encoder and decoder. Cai, et al.
\cite{CaiContextTreeSideInfo} have shown how to modify the context-tree method to account for
side information at the encoder. It is also possible to modify the Lempel-Ziv algorithms to
account for side information at the encoder (\cite{LempelZivSideInfo}, \cite{BergerSideInfo}).

    The purpose of this paper, however, is to consider how to compress when the side
information is available to the decoder only. This restriction disallows variable length codes
as a generic solution. Variable length codes work because they assign short codewords to
typical source strings and longer codewords to atypical strings. When the side information is
available only to the decoder, the encoder cannot tell when the joint source is behaving
atypically. As an example consider a binary equiprobable source $U$. Let $V$ be the output of
$U$ passed through a binary symmetric channel with crossover probability $1/10$. Every $U$
source string of the same length occurs with equal probability, but clearly the side
information allows the source to be compressed below $1$ bit per symbol.

    One approach around this problem is to use block codes such as LDPC codes to give a
`structured' binning of the source strings. The side information is then used at the decoder to
distinguish amongst the source strings in the received bin. In the same mold, it is also
possible to use turbo-codes as done by Aaron, Girod, et.al. (\cite{AaronTurbo},
\cite{GirodVideoCoding}). Regardless of the type of code, lack of the side information at the
encoder somehow necessitates a shift in `complexity' from the encoder to the decoder.

    The idea of shifting complexity from encoder to decoder in lossless source coding is not
new. In \cite{Hellman}, Hellman suggested using convolutional codes for joint source-channel
coding in applications such as deep-space communications where computational effort at the
encoder comes at a premium. Around the same time, papers of Koshelev \cite{Koshelev} and
Blizard \cite{Blizard} suggested using convolutional codes in conjunction with sequential
decoders for the purposes of data compression and joint source-channel coding. These ideas
extend naturally to the subject of this paper, lossless source coding and joint source-channel
coding with side information available to the decoder only.

    The approach of this paper is to use random, time-varying, infinite constraint length
convolutional codes to sequentially `bin' an IID source and a Stack Algorithm sequential
decoder to (almost) losslessly recover it. The decoder has a variable `bias' parameter, as in
\cite{JelinekUpperBound} by Jelinek, that allows for a tradeoff between probability of error
and moments of the random variable of computation associated with the sequential decoder. The
proof techniques are adaptations to source coding and joint source-channel coding of those of
\cite{JelinekUpperBound}.

\begin{table*}
\begin{center}
\begin{tabular}{|c|c|c|c|c|}
  \hline
   & Channel Coding & Source Coding & Source Coding with SI & Joint S-C Coding with SI\\
  \hline
 Block Codes & Shannon \cite{ShannonWeaver} & Shannon \cite{ShannonWeaver} & Slepian and Wolf \cite{SlepianWolf} & - \\ \hline
 Block Error Exponent & Gallager \cite{GallagerBook} & Csiszar and K\"{o}rner \cite{CsiszarBook}& Gallager \cite{GallagerSideInformation} & - \\ \hline
 Convolutional and & Elias \cite{EliasConvCoding} &  Hellman \cite{Hellman}, & - & - \\
  Tree Codes& & Blizard \cite{Blizard} & &\\ \hline
 Delay Error Exponent & Pinsker \cite{Pinsker}, Sahai \cite{SahaiDelay} & Chang and Sahai \cite{ChangLossless} & Chang and
 Sahai \cite{ChangISIT06} & Chang and Sahai \cite{ChangAllerton06} \\
 \hline
 Sequential Decoding  & Jelinek \cite{JelinekUpperBound}& Thm. \ref{thm:proberrorsi}& Thm. \ref{thm:proberrorsi} & Thm. \ref{thm:proberrorsc} \\
 Delay Error Exponent  & & & & \\ \hline
 Sequential Decoding & Jelinek \cite{JelinekComp} & Koshelev \cite{Koshelev}& Thm. \ref{thm:compsi} & Thm. \ref{thm:compsc} \\
 Computation Achievability & Savage \cite{Savage}& & & \\ \hline
 Sequential Decoding  & Jacobs and  & Arikan and & open & open \\
Computation Converse & Berlekamp \cite{JacobsBerlekamp} & Merhav  \cite{ArikanSourceChannel}& &
\\  \hline
\end{tabular}
 \caption{Some references in data compression and channel coding and data
compression with side information.} \label{table:historytable}
\end{center}
\end{table*}

    Table \ref{table:historytable} shows the relation of this paper with some prior work. There
are several lines of work in information theory that our scheme is related to. As already
mentioned, the main point of this paper is to extend the idea of using convolutional encoding
with sequential decoding for lossless source coding by modifying the decoder to allow the use
of side information.

    In \cite{Koshelev}, Koshelev shows that there is a point-to-point source coding `cutoff rate' for a stack-based sequential decoding algorithm. That is, if the rate is larger than the cutoff rate, then
the expected mean of computation performed by the sequential decoder is finite. Work in the
opposite direction by Arikan and Merhav \cite{ArikanSourceChannel} showed that this cutoff rate
is tight; if the rate is below the cutoff rate, the expected mean in computation is infinite.
Furthermore, \cite{ArikanSourceChannel} gives a lower bound to the cutoff rate for all moments
of the random variable of computation, not only the mean. Our result regarding computation
parallels Koshelev's, only with side information allowed at the decoder. We give an upper bound
to the `cutoff rate' for moments in the interval $[0,1]$, of sequential decoding for lossless
source coding with side information at the decoder. When the side information is independent of
the source to be recovered, reducing to the point-to-point version of the problem, this cutoff
rate coincides with that of \cite{ArikanSourceChannel}.

    One interesting aspect of our scheme is its `anytime' or delay-universal nature. By using
an infinite constraint-length convolutional code, it is possible to have a probability of error
that goes to zero exponentially with delay\footnote{Delay is defined as the difference between
the decoding time and the time the symbol entered the encoder.}. For certain problems in
distributed control (\cite{SahaiPartI}, \cite{Allerton05paper}), an exponentially decreasing
probability of error is required to guarantee plant stability (in a moment sense). For these
problems, the error exponent with delay determines the moments of the plant state that can be
stabilized. The scheme presented in this paper, if there is a channel between encoder and the
side-information aided decoder, achieves an error exponent with delay analogous to the
point-to-point random block coding error exponent of Problem 5.16 of Gallager
\cite{GallagerBook}. Recent work by Chang, et. al. (\cite{ChangISIT05}, \cite{ChangLossless},
\cite{ChangISIT06}, \cite{ChangCISS07}) has shown that in general, the best block error
exponents are much lower than the best error exponents with delay achievable for problems of
lossless source coding with and without side information.

    The paper is organized as follows. In Section \ref{sec:probdef} we set up the problem of
streaming source coding, perhaps with a noisy channel between encoder and decoder, and with
side information available to the decoder. Then in Section \ref{sec:binningscheme} we give a
description of the encoder and decoder. Next, in Section \ref{sec:mainresults} we state the two
main theorems One theorem states the error exponent with delay for this scheme and the other
theorem gives an upper bound to the asymptotic distribution of computation when using the stack
algorithm. The next section gives some examples and simulation results showing that the
proposed scheme can be implemented with non-prohibitive complexity. In the conclusion, we
discuss some open questions left in this specific line of work and some future directions.
Finally, in the appendix, we give proofs of the theorems of the text.

\section{Sequential data compression with side information}
\subsection{Problem definition}\label{sec:probdef} The source is modelled as a sequence of IID random variables $(U_i,V_i),~ i \geq 1,$ that take on values from a finite alphabet
$\mU \times \mV$. Each $(U_i,V_i)$ is drawn according to a probability mass function $Q(u,v)$.
With some abuse of notation, we will use $Q(u)$, for $u \in \mU$, to denote the marginal
probability $\sum_{v\in \mV} Q(u,v)$. Similarly, $Q(v)$ will be the marginal $\sum_{u\in \mU}
Q(u,v)$ for $v \in \mV$. Finally, $Q(u|v) = Q(u,v)/Q(v)$ for $(u,v) \in \mU \times \mV$.
Without loss of generality, assume $Q(v) > 0$, $\forall ~v \in \mV$. If $U$ and $V$ are
independent, the point-to-point source coding problem is recovered.

Our goal is to code the $U_i$ symbols causally into a fixed rate bit stream so that the symbols
can be recovered losslessly by a decoder in the sense that a symbol $U_i$ is recovered with
probability $1$ in the limit of large decoding delay. For reasons mentioned in the
introduction, a truly fixed rate coding strategy that assigns the same number of bits to
sequences of the same length will be pursued.

\vspace{.1in}

\begin{figure}[htb]
\begin{center}
 \setlength{\unitlength}{1.5mm}
\begin{picture}(140,30)(-15,0)
\normalsize
 \put(20, 30){$u_1$} \put(30, 30){$u_2$} \put(40,
30){$u_3$} \put(50, 30){$u_4$} \put(60, 30){$u_5$} \put(70, 30){$u_6$ ...}

  \put(30, 20){$B_1(u_1^2)$}
  \put(50, 20){$B_2(u_1^4)$}
  \put(70, 20){$B_3(u_1^6)$ ...}

\put(20, 10){$\widehat{u}_1^1(1)$} \put(30, 10){$\widehat{u}_1^2(2)$}  \put(40,
10){$\widehat{u}_1^3(3)$} \put(50, 10){$\widehat{u}_1^4(4)$} \put(60, 10){$\widehat{u}_1^5(5)$}
\put(70, 10){$\widehat{u}_1^6(6)$ ...}

\put(20,  0){$v_1$} \put(30,  0){$v_2$} \put(40,  0){$v_3$} \put(50,  0){$v_4$} \put(60,
0){$v_5$} \put(70,  0){$v_6$ ...} \put(0, 10){Decoding}

\put(0, 20){Encoding}

\put(0, 30){Source}

\put(0, 0){Side-info} \multiput(21,29)(10,0){6}{\vector(0,-1){5}}
\multiput(31,19)(20,0){3}{\vector(0,-1){5}}
 \multiput(21,4)(10,0){6}{\vector(0,1){5}}
    \end{picture}
    \caption{ Sequential source coding with side-information:
    rate $R=\frac{1}{2}$.}
    \label{fig:time_line}
    \end{center}
\end{figure}
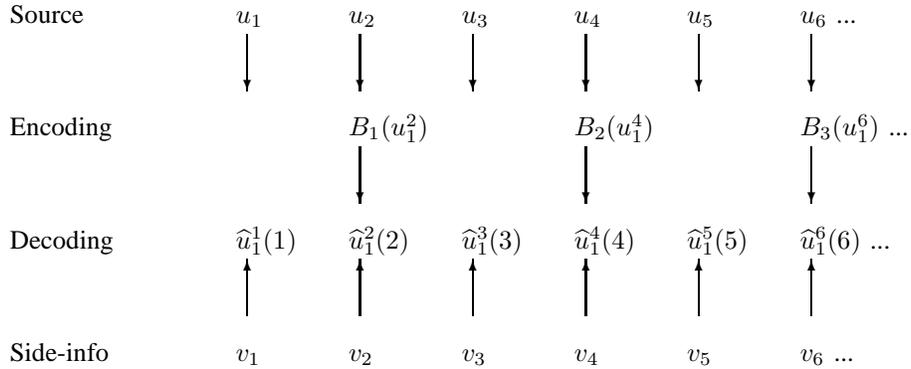

Figure \ref{fig:time_line} shows the setup of our `streaming' source coding problem. At a
discrete time instant $n$, the encoder has access to the source realization up through time
$n$, which is denoted\footnote{We will use $z_i^j$ to denote the vector $(z_i, z_{i+1},\ldots,
z_j)$ if $i \leq j$ and the null string if $i > j$.} $u_1^n$. Let the rate of the encoder be
$R$ bits per source symbol. The encoder at time $n$ outputs $\lfloor nR \rfloor - \lfloor
(n-1)R \rfloor$ bits that are a function of $u_1^n$. Based on the bits $B_1^{\lfloor nR
\rfloor}$ and the side information $v_1^n$, the decoder at time $n$ gives its estimate of the
source symbols up through time $n$, denoted as $\widehat{u}_1^{n}(n)$.
\begin{eqnarray}
    \mE_n & : & \mU^n \rightarrow \{0,1\}^{\lfloor nR \rfloor - \lfloor (n-1)R \rfloor }\\
    B_{\lfloor (n-1)R \rfloor}^{\lfloor nR \rfloor}(u_1^n) & = & \mE_n (u_1^n) \\
    \mD_n & : & \{0,1\}^{\lfloor nR \rfloor} \times \mV^n \rightarrow \mU^n \\
    \widehat{u}_1^n (n) & = & \mD_n(B_1^{\lfloor nR \rfloor}, v_1^n)
\end{eqnarray}

The only interesting values of $R$ lie in the interval $[H(U|V), \log_2(|\mU|)]$ since we need
a rate of at least the conditional entropy to losslessly encode the source, and if $R
> \log_2(|\mU|)$, we could just index the source sequences on a per-letter basis and losslessly recover them
with no delay.
    \begin{equation}
        H(U|V) \triangleq \sum_{v \in \mV} Q(v)\sum_{u \in \mU} Q(u|v) \log_2 \frac{1}{Q(u|v)}
    \end{equation}

There are two measures of performance that we will evaluate. First is the tradeoff between
probability of error and delay.
\begin{definition} The {\em probability of error with delay $d$}, $P_e(d)$, is
\label{def:sourceerror}
\begin{equation} P_e(d) \triangleq \sup_{n} P(\hat{u}_1^n(n+d) \neq u_1^n) \end{equation}
This probability is taken over the randomness in the source and any randomness that may be
present in the encoder or decoder. The {\em error exponent with delay}, or reliability exponent
$E(R)$ at the rate $R$ where the encoder/decoder operates is
\begin{equation} E(R) \triangleq \liminf_{d \rightarrow \infty} -\frac{1}{d}\log_2 P_e(d) \end{equation}
\end{definition} \vspace{.1in}

The second measure of performance lies in the random variable of computation. The motivation
for developing sequential decoders has always been the opportunity to have a `nearly optimal'
decoder without exponentially growing complexity in block length or delay \cite{ForneySeq}. The
amount of computation performed by our source decoder will be measured in the number of source
sequences that are considered or compared against others.

\begin{definition} If $u_n$ is the true source realization at time $n \geq 1$, the {\em $i^{th}$ incorrect
subtree}, $\mC_i$, is
\begin{equation} \mC_i \triangleq \big \{ z_1^n \in \mU^n: n \geq i, \ z_1^{i-1} =u_1^{i-1} \textrm{ and } z_i \neq u_i\big \}
\end{equation}
The {\em $i^{th}$ random variable of computation}, $N_i$, is the number of nodes in $\mC_i$
that are ever examined by the decoder.
\end{definition} \vspace{.1in}

The definition of $N_i$ is a bit vague for arbitrary decoders but becomes concrete for
sequential decoders, because the defining property of sequential decoders is essentially that
they examine paths in a tree or trellis structure one by one.\footnote{There is also some
amount of `internal' computation the decoder must do to determine the codewords of the source
sequences. We assume an oracle gives the decoder any source codeword it wants at unit cost.
This is somewhat significant in our random convolutional code implementation since the
encoder's output depends on all previous source symbols. This means that as time increases,
there is an increasing complexity to determining the bits assigned to a source symbol.}

\subsection{A random binning scheme with a stack decoder}\label{sec:binningscheme}

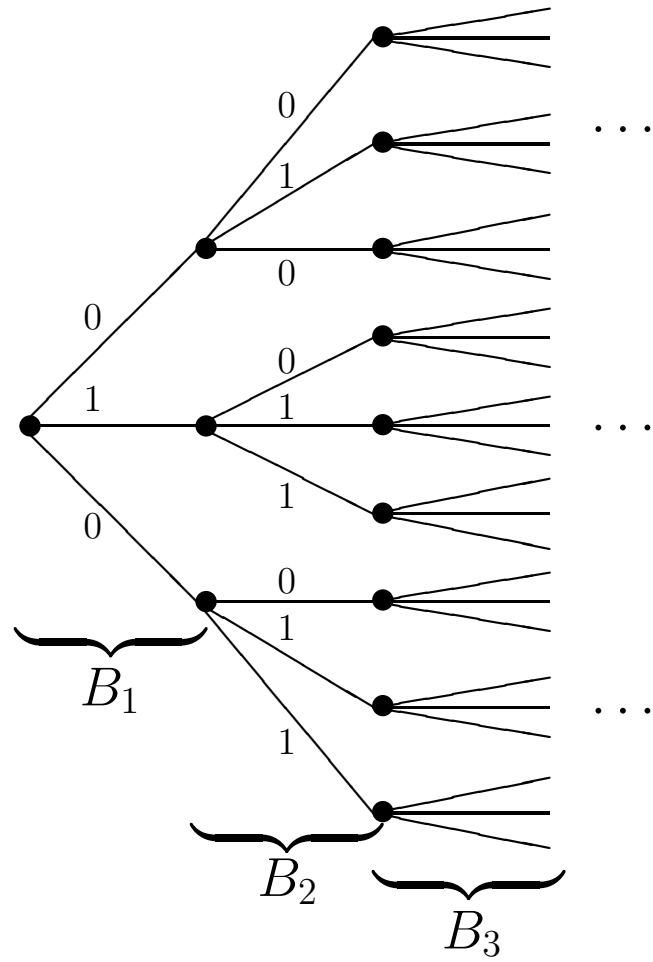
\begin{figure}[hbp]
\begin{center}

\setlength{\unitlength}{3700sp}%
\begingroup\makeatletter\ifx\SetFigFont\undefined%
\gdef\SetFigFont#1#2#3#4#5{%
  \reset@font\fontsize{#1}{#2pt}%
  \fontfamily{#3}\fontseries{#4}\fontshape{#5}%
  \selectfont}%
\fi\endgroup%
\begin{picture}(3795,6300)(1160,-5349)
\thicklines \put(1182,-1996){\line( 1, 0){1181}} \put(1182,-1996){\line( 1, 1){1181}}
\put(1182,-1996){\line( 1,-1){1181}}
\put(1150,-2090){\makebox(0,0)[lb]{\smash{{\SetFigFont{20}{24.0}{\rmdefault}{\mddefault}{\updefault}$\bullet$}}}}

\put(2363,-1996){\line( 2, 1){1182}}\put(2363,-1996){\line( 2,-1){1182}}
\put(2422,-1996){\line( 1, 0){1122}}
\put(2330,-2090){\makebox(0,0)[lb]{\smash{{\SetFigFont{20}{24.0}{\rmdefault}{\mddefault}{\updefault}$\bullet$}}}}

\put(2363,-815){\line( 5, 6){1180}} \put(2363,-815){\line( 5, 3){1180}} \put(2363,-815){\line(
1, 0){1181}}
\put(2330,-900){\makebox(0,0)[lb]{\smash{{\SetFigFont{20}{24.0}{\rmdefault}{\mddefault}{\updefault}$\bullet$}}}}

\put(2363,-3177){\line( 1, 0){1181}} \put(2363,-3177){\line( 5,-3){1180}}
\put(2363,-3177){\line( 5,-6){1180}}
\put(2330,-3267){\makebox(0,0)[lb]{\smash{{\SetFigFont{20}{24.0}{\rmdefault}{\mddefault}{\updefault}$\bullet$}}}}

\put(3544,603){\line( 6, 1){1182}} \put(3544,603){\line( 1, 0){1181}} \put(3544,603){\line(
6,-1){1182}}
\put(3514,520){\makebox(0,0)[lb]{\smash{{\SetFigFont{20}{24.0}{\rmdefault}{\mddefault}{\updefault}$\bullet$}}}}

\put(3544,-106){\line( 6, 1){1182}} \put(3544,-106){\line( 1, 0){1181}} \put(3544,-106){\line(
6,-1){1182}}
\put(3514,-186){\makebox(0,0)[lb]{\smash{{\SetFigFont{20}{24.0}{\rmdefault}{\mddefault}{\updefault}$\bullet$}}}}

\put(3544,-815){\line( 5, 1){1180}} \put(3544,-815){\line( 1, 0){1181}} \put(3544,-815){\line(
6,-1){1182}}
\put(3514,-900){\makebox(0,0)[lb]{\smash{{\SetFigFont{20}{24.0}{\rmdefault}{\mddefault}{\updefault}$\bullet$}}}}

\put(3544,-1405){\line( 6, 1){1182}} \put(3544,-1405){\line( 1, 0){1181}}
\put(3544,-1405){\line( 6,-1){1182}}
\put(3514,-1485){\makebox(0,0)[lb]{\smash{{\SetFigFont{20}{24.0}{\rmdefault}{\mddefault}{\updefault}$\bullet$}}}}

\put(3544,-1996){\line( 6, 1){1182}} \put(3544,-1996){\line( 1, 0){1181}}
\put(3544,-1996){\line( 6,-1){1182}}
\put(3514,-2080){\makebox(0,0)[lb]{\smash{{\SetFigFont{20}{24.0}{\rmdefault}{\mddefault}{\updefault}$\bullet$}}}}

\put(3544,-2586){\line( 5, 1){1180}} \put(3544,-2586){\line( 1, 0){1181}}
\put(3544,-2586){\line( 5,-1){1180}}
\put(3514,-2675){\makebox(0,0)[lb]{\smash{{\SetFigFont{20}{24.0}{\rmdefault}{\mddefault}{\updefault}$\bullet$}}}}

\put(3544,-3177){\line( 6, 1){1182}} \put(3544,-3177){\line( 1, 0){1181}}
\put(3544,-3177){\line( 6,-1){1182}}
\put(3514,-3257){\makebox(0,0)[lb]{\smash{{\SetFigFont{20}{24.0}{\rmdefault}{\mddefault}{\updefault}$\bullet$}}}}

\put(3544,-3885){\line( 1, 0){1181}} \put(3544,-3885){\line( 6, 1){1182}}
\put(3544,-3885){\line( 6,-1){1182}}
\put(3514,-3965){\makebox(0,0)[lb]{\smash{{\SetFigFont{20}{24.0}{\rmdefault}{\mddefault}{\updefault}$\bullet$}}}}

\put(3544,-4594){\line( 1, 0){1181}} \put(3544,-4594){\line( 5, 1){1180}}
\put(3544,-4594){\line( 5,-1){1180}}
\put(3514,-4675){\makebox(0,0)[lb]{\smash{{\SetFigFont{20}{24.0}{\rmdefault}{\mddefault}{\updefault}$\bullet$}}}}

\put(1600,-1350){\makebox(0,0)[lb]{\smash{{\SetFigFont{14}{16.8}{\rmdefault}{\mddefault}{\updefault}$0$}}}}
\put(1600,-1900){\makebox(0,0)[lb]{\smash{{\SetFigFont{14}{16.8}{\rmdefault}{\mddefault}{\updefault}$1$}}}}
\put(1600,-2750){\makebox(0,0)[lb]{\smash{{\SetFigFont{14}{16.8}{\rmdefault}{\mddefault}{\updefault}$0$}}}}
\put(2900,-1641){\makebox(0,0)[lb]{\smash{{\SetFigFont{14}{16.8}{\rmdefault}{\mddefault}{\updefault}$0$}}}}
\put(2900,-1950){\makebox(0,0)[lb]{\smash{{\SetFigFont{14}{16.8}{\rmdefault}{\mddefault}{\updefault}$1$}}}}
\put(2900,-2550){\makebox(0,0)[lb]{\smash{{\SetFigFont{14}{16.8}{\rmdefault}{\mddefault}{\updefault}$1$}}}}
\put(2900,-1051){\makebox(0,0)[lb]{\smash{{\SetFigFont{14}{16.8}{\rmdefault}{\mddefault}{\updefault}$0$}}}}
\put(2900,
71){\makebox(0,0)[lb]{\smash{{\SetFigFont{14}{16.8}{\rmdefault}{\mddefault}{\updefault}$0$}}}}
\put(2900,-400){\makebox(0,0)[lb]{\smash{{\SetFigFont{14}{16.8}{\rmdefault}{\mddefault}{\updefault}$1$}}}}
\put(2900,-3118){\makebox(0,0)[lb]{\smash{{\SetFigFont{14}{16.8}{\rmdefault}{\mddefault}{\updefault}$0$}}}}
\put(2900,-4200){\makebox(0,0)[lb]{\smash{{\SetFigFont{14}{16.8}{\rmdefault}{\mddefault}{\updefault}$1$}}}}
\put(2900,-3431){\makebox(0,0)[lb]{\smash{{\SetFigFont{14}{16.8}{\rmdefault}{\mddefault}{\updefault}$1$}}}}
\put(1141,-3372){\makebox(0,0)[lb]{\smash{{\SetFigFont{20}{24.0}{\rmdefault}{\mddefault}{\updefault}$\underbrace{\hspace{1in}}$}}}}
\put(3540,-5000){\makebox(0,0)[lb]{\smash{{\SetFigFont{20}{24.0}{\rmdefault}{\mddefault}{\updefault}$\underbrace{\hspace{1in}}$}}}}
\put(2322,-4685){\makebox(0,0)[lb]{\smash{{\SetFigFont{20}{24.0}{\rmdefault}{\mddefault}{\updefault}$\underbrace{\hspace{1in}}$}}}}
\put(1575,-3870){\makebox(0,0)[lb]{\smash{{\SetFigFont{20}{24.0}{\rmdefault}{\mddefault}{\updefault}$B_1$}}}}
\put(2765,-5150){\makebox(0,0)[lb]{\smash{{\SetFigFont{20}{24.0}{\rmdefault}{\mddefault}{\updefault}$B_2$}}}}
\put(4000,-5500){\makebox(0,0)[lb]{\smash{{\SetFigFont{20}{24.0}{\rmdefault}{\mddefault}{\updefault}$B_3$}}}}
\put(5000,-2100){\makebox(0,0)[lb]{\smash{{\SetFigFont{20}{24.0}{\rmdefault}{\mddefault}{\updefault}$\cdots$}}}}
\put(5000,-100){\makebox(0,0)[lb]{\smash{{\SetFigFont{20}{24.0}{\rmdefault}{\mddefault}{\updefault}$\cdots$}}}}
\put(5000,-4000){\makebox(0,0)[lb]{\smash{{\SetFigFont{20}{24.0}{\rmdefault}{\mddefault}{\updefault}$\cdots$}}}}
\end{picture}%
\end{center}
\caption[A ternary tree]{An example of a tree code for a source with ternary alphabet. Here the
rate $R$ is one bit per source symbol.} \label{fig:ppsourcetreecode}
\end{figure}

In this section, the encoder and decoder for the coding strategy of this paper is described.
The encoder used is similar to the encoder used in the sequential source coding paper of
\cite{ChangISIT05}. The bit sequence is arrived at by the use of a random tree code, which can
be implemented using a time-varying, infinite constraint length, random convolutional code.
Figure \ref{fig:ppsourcetreecode} shows an example of such a code. We first envisage a uniform
tree with $|\mathcal{U}|$ branches emanating from each node. The branches are numbered $1,2,
\ldots, |\mathcal{U}|$ to denote the extension of the parent sequence by one symbol from
$\mathcal{U}$. Hence, for all $k \geq 1$ there is a one-to-one correspondence between
$|\mathcal{U}|$-ary strings of length $k$ and nodes in the tree. These properties make clear
that labelling the branches of the tree with an appropriate number of code bits would yield a
tree encoding of the source: a sequential source code.

The sequential random binning scheme we use is an ensemble of tree codes, with every bit on
every branch drawn identically and independently as Bernoulli $(1/2,1/2)$ ($\mB(1/2)$) random
variables. This means that if source sequences $u_1^n$ and $z_1^n$ are the same until time
$n-d+1$, i.e. $u_1^{n-d} = z_1^{n-d}$, but $u_{n-d+1} \neq z_{n-d+1}$, the probability that
$u_1^n$ and $z_1^n$ are placed in the same `bin' is $2^{-dR}$. This is because the last $dR$
bits of the codewords for $u_1^n$ and $z_1^n$ are drawn IID $\mB(1/2)$. We refer to the bits in
the codewords of source sequences as `parities' because we think of them as coming from a
time-varying, infinite constraint length convolutional code.

Decoding will be done by a stack algorithm, and hence is also sequential. For explanations of
the stack algorithm, refer to \cite{ZigangirovBook}, \cite{LinCostelloBook}, or
\cite{WickerBook}. The following is the specific stack algorithm used. We initialize the stack
with the root node having a metric of $0$.

\begin{enumerate}
\item Let $u_1^l$ denote the (partial) source sequence at the top of the stack. Remove $u_1^l$
from the stack and consider each of its $|\mU|$ extensions by one symbol from $\mU$, i.e.
$(u_1^l, u)$, $\forall u \in \mX$. Let $\tilde{u}_1^{l+1}$ be one of these extensions, and do
the following for each of the extensions. If the parities of $\tilde{u}_1^{l+1}$ match the
parities received, update the metric of $\tilde{u}_1^{l+1}$ and add it to the stack in a sorted
way (highest metric on top). Otherwise discard $\tilde{u}_1^{l+1}$. \footnote{ Note that the
parities of $\tilde{u}_1^{l+1}$ will match those received if and only if the label on the
branch extending $u_1^l$ to $\tilde{u}_1^{l+1}$ matches the $R$ parities received in the last
time step.}

\item Let $u_1^k$ denote the sequence on top of the stack after all the relevant extensions
have been added. If the length of $u_1^k$, $k$, is up to the current time, declare $u_1^k$ as
the decoded source sequence so far. Otherwise repeat 1.
\end{enumerate}
The metrics are updated in an additive manner, with the metric of $\tilde{u}_1^{l+1}$ being
$\Gamma(\tilde{u}_1^{l+1}) = \Gamma(\tilde{u}_1^l) + \Gamma(\tilde{u}_{l+1})$. For $(u,v) \in
\mU\times \mV$, the metric for the source symbol $u$ given side information $v$ is
\begin{displaymath}
\Gamma(u) \triangleq G + \log_2(Q(u|v))
\end{displaymath}

The parameter $G$ is the `bias' and controls to a large extent the amount of searching through
the tree the algorithm performs. The bias is used as a normalizer so that the true path through
the tree has a metric that is slowly increasing in time, while false path metrics are dropped
to $- \infty$ by non-matching parities.


\subsection{Joint source-channel coding with side information} \label{sec:jointscwithsisetup}

Suppose there is a DMC between the encoder and the decoder. Let $W$ be its probability
transition matrix, from a finite input alphabet $\mX$ to a finite output alphabet $\mY$. Assume
there are $\lambda > 0$ channel uses per source symbol.
    \begin{eqnarray}
        \mE_n & : & \mU^n \rightarrow \mX^{\lfloor n\lambda \rfloor - \lfloor (n-1)\lambda \rfloor }\\
        x_{\lfloor (n-1)\lambda \rfloor}^{\lfloor n\lambda \rfloor}(u_1^n) & = & \mE_n (u_1^n) \\
        \mD_n & : & \mY^{\lfloor n\lambda \rfloor} \times \mV^n \rightarrow \mU^n \\
        \widehat{u}_1^n (n) & = & \mD_n(y_1^{\lfloor n\lambda \rfloor}, v_1^n)
    \end{eqnarray}

The random binning encoder and the stack decoder of the previous section changes only slightly.
First, the encoding tree is restricted to having one channel symbol on each branch, rather than
$R$ bits. We will assume each channel input on the tree is drawn IID from a distribution
$\beta(x)$ on $\mX$. Secondly, the stack decoder cannot discard paths based on parities
anymore. So, if $u$,$v$,$x_1^\lambda$, and $y_1^\lambda$ are respectively the source symbol on
a branch, side information symbol, channel inputs on the branch and the channel outputs
received by the decoder, then the decoder assigns a metric of:
    \begin{displaymath} \Gamma(u) = G + \log_2\frac{Q(u|v)W(y_1^\lambda|x_1^\lambda)}{P(y_1^\lambda)} \end{displaymath}

where $P(y) \triangleq \sum_{x\in\mX} \beta(x)W(y|x)$ and $P(y_1^\lambda) = \prod_{k=1}^\lambda
P(y_k)$. The performance measures remain the same, with the error exponent at `rate' $\lambda$
being $E(\lambda) = \liminf_{d\rightarrow \infty} -\frac{1}{d}\log_2 P_e(d)$.

\clearpage

\section{Main Results} \label{sec:mainresults}
\subsection{Functions of interest} \label{sec:gallagerfunctions}
We start with the definition of some functions that appear in the theorem statements. The
following functions of the channel input distribution, $\beta(x)$, and channel transition
probability matrix, $W(y|x)$, appear in \cite{JelinekUpperBound}.
    \begin{eqnarray}
        E_0(\rho) & \triangleq & - \log_2 \B[ \sum_{y\in \mY} P(y) \b( \sum_{x\in\mX}
        \beta(x)\b(\frac{W(y|x)}{P(y)}\b)^\frac{1}{1+\rho} \b)^{1+\rho} \B] \\
        F(\rho) & \triangleq & - \log_2 \B[ \sum_{y\in \mY} P(y) \b( \sum_{x\in\mX}
        \beta(x)\b(\frac{W(y|x)}{P(y)}\b)^\frac{1}{1+\rho} \b)^\rho \B] \\
        G(\rho) & \triangleq & - \log_2 \B[ \sum_{y\in \mY} P(y) \b( \sum_{x\in\mX}
        \beta(x)\b(\frac{W(y|x)}{P(y)}\b)^\frac{1}{1+\rho} \b)\B]
\end{eqnarray}

We define the following functions of the source distribution for $\rho \geq 0$. $E_{si}(\rho)$
can be found in \cite{GallagerSideInformation} and the others are modifications of $E_{si}$.
    \begin{eqnarray}
        E_{si}(\rho) & \triangleq   & \log_2 \B[ \sum_{v\in \mV} Q(v) \b( \sum_{u\in \mU} Q(u|v)^{\frac{1}{1+\rho}} \b)^{1+\rho}\B] \\
        F_{si}(\rho) & \triangleq & \log_2\B[ \sum_{v \in \mV} Q(v) \b ( \sum_{u\in \mU}
        Q(u|v)^{\frac{1}{1+\rho}} \b )^{\rho} \B]\\
        G_{si}(\rho) & \triangleq & \log_2 \B[ \sum_{v \in \mV} Q(v) \b ( \sum_{u\in \mU}
        Q(u|v)^{\frac{1}{1+\rho}} \b )\B]
    \end{eqnarray}

If the side information is independent of $U$, then we get the simpler functions $E_s(\rho)$,
$F_s(\rho)$ and $G_s(\rho)$ below.
    \begin{eqnarray}
        E_s(\rho) & \triangleq & (1+\rho) \log_2 \B[ \sum_{u\in \mU} Q(u)^\frac{1}{1+\rho} \B] \\
        F_s(\rho) & \triangleq & \rho \log_2 \B[ \sum_{u\in \mU} Q(u)^\frac{1}{1+\rho} \B] \\
        G_s(\rho) & \triangleq & \log_2 \B[ \sum_{u\in \mU} Q(u)^\frac{1}{1+\rho} \B]
    \end{eqnarray}


\subsection{Probability of error with delay}
\begin{theorem}[Error exponent with delay for source coding with side
    information]\label{thm:proberrorsi} Suppose that the decoder has access to the side information
    and there is a noiseless rate $R$ binary channel between the encoder and decoder. Fix any $\eps
    > 0$ and let $\rho \in [0,1]$. For the encoder/decoder of Section \ref{sec:binningscheme}, if the bias $G$ satisfies
    \begin{equation} G \leq \frac{1+\rho}{\rho}\B[ E_{si}(\rho) - F_{si}(\rho) \B]
    \end{equation}
    then there is a constant $K_\eps < \infty$ so that
    \begin{equation} P_e(d) < K_\eps \exp_2 \B(-d\b(\rho R - E_{si}(\rho) - \eps\b)\B)
    \end{equation}
    Hence, with suitable choice of bias, the error exponent with delay can approach
    \begin{equation} E(R) = E_{r,si}(R) \triangleq \sup_{\rho \in [0,1]} \rho R - E_{si}(\rho)
    \end{equation}
\end{theorem}\vspace{.2in}

If the side information is independent of the source $U$, then $E_{si}$, $F_{si}$ and $G_{si}$
simplify to $E_s$, $F_s$ and $G_s$ respectively. So, in the case of straight point-to-point
lossless source coding, we arrive at an source coding equivalent of Gallager's random coding
exponent:
    \begin{equation} E_{r,pp}(R) \triangleq \max_{\rho \in [0,1]} \rho R - E_s(\rho)
    \end{equation}

\begin{theorem}[Error exponent with delay for joint source-channel coding with side information] \label{thm:proberrorsc} Suppose there is a channel $W$ between the encoder and the decoder and
side information is available to the decoder. Fix any $\eps > 0$ and let $\rho \in [0,1]$. For
the encoder/decoder of Sections \ref{sec:binningscheme} and \ref{sec:jointscwithsisetup}, if
the bias $G$ satisfies
\begin{equation} G \leq \frac{1+\rho}{\rho}\B[ E_{si}(\rho) - F_{si}(\rho) - \lambda E_0(\rho) + \lambda F(\rho) \B]
\end{equation}
then, there is a constant $K_\eps < \infty$ so that
\begin{equation} P_e(d) < K_\eps \exp_2 \B(-d\b(\lambda E_0(\rho) - E_{si}(\rho) - \eps \b)\B)
\end{equation}
Hence, with suitable choice of bias, the error exponent with delay can be
\begin{equation} E(\lambda) = E_{r,jscsi}(\lambda) \triangleq \sup_{\rho \in [0,1]} \lambda E_0(\rho) - E_{si}(\rho)
\end{equation} \end{theorem}\vspace{.2in}

By assuming the side information to be independent of the source, we once again have a scheme
for joint source-channel coding. The error exponent achieved is the joint source-channel
equivalent of Gallager's random coding exponent (\cite{GallagerBook}, Problem 5.16).
    \begin{equation} E_{r, jsc}(\lambda) \triangleq \max_{\rho \in [0,1]} \lambda E_0(\rho) - E_s(\rho)
    \label{eqn:jscexponent}
    \end{equation}

The exponent of (\ref{eqn:jscexponent}) is lower in general than the joint source-channel
exponent of Csiszar \cite{CsiszarJSCExponent}.

\subsection{Random variable of computation}

\begin{theorem}[Computation of stack decoder with side information] \label{thm:compsi} Suppose that the decoder has access to the side
information and there is a rate $R$ noiseless, binary channel between the encoder and decoder.
Fix any $\gamma \in [0,1]$. For the encoder/decoder of Section \ref{sec:binningscheme}, if the
bias $G$ satisfies
\begin{equation} \frac{1+\gamma}{\gamma}G_{si}(\gamma) < G <
\frac{1+\gamma}{\gamma} \B[ \gamma R - F_{si}(\gamma)\B] \label{eqn:biassi} \end{equation} then
the $\gamma^{th}$ moment of computation is uniformly finite all for $i$, i.e. $\exists ~K <
\infty$ such that $\forall~ i,~ E[N_i^\gamma] < K$, if
\begin{equation}
R > \frac{E_{si}(\gamma)}{\gamma} \end{equation} \end{theorem} \vspace{.2in}

As a conclusion of the theorem, we show that the interval of viable bias values implicit in
\ref{eqn:biassi} is in fact non-empty if $ R > E_{si}(\gamma)/\gamma$.


By restricting to the point-to-point case, we see that the $\gamma^{th}$ moment of computation,
for $\gamma \in [0,1]$, can be finite if
    \begin{equation} R > \frac{E_s(\gamma)}{\gamma} \end{equation}
This result has been known for $\gamma = 1$, due to Koshelev \cite{Koshelev}. We conjecture
that Theorem \ref{thm:compsi} remains true for $\gamma > 1$. This conjecture is supported by
simulation, but unproven. It is established by using the results found in
\cite{ArikanSourceChannel} that if $R < E_s(\gamma)/\gamma$, then $E[N_i^\gamma]$ cannot be
uniformly bounded. Together, these results tell us that our stack decoder is doing as well as
could be hoped for any sequential decoder in terms of the moments of computation for the
point-to-point case.

\begin{theorem}[Computation of stack decoder for joint source-channel coding with side information] \label{thm:compsc} Suppose there is a channel $W$ between the encoder and the decoder and
side information is available to the decoder. Fix any $\gamma \in [0,1]$. For the
encoder/decoder of Sections \ref{sec:binningscheme} and \ref{sec:jointscwithsisetup}, if the
bias $G$ satisfies
\begin{equation} \frac{1+\gamma}{\gamma}\B[ G_{si}(\gamma) - \lambda G(\gamma) \B] < G <
\frac{1+\gamma}{\gamma} \B[\lambda F(\gamma) - F_{si}(\gamma)\B] \label{eqn:biassc}
\end{equation} then the $\gamma^{th}$ moment of computation is uniformly finite all for $i$,
i.e. $\exists ~K < \infty$ such that $\forall~ i,~ E[N_i^\gamma] < K$, if
\begin{equation}
\lambda E_0(\gamma) > E_{si}(\gamma)\end{equation} \end{theorem}\vspace{.2in}

Again, in the appendix, we show that if $E_0(\gamma) > E_{si}(\gamma)$, then the interval of
acceptable bias values in \ref{eqn:biassc} is non-empty.

By removing the side information, we see the condition needed for a finite $\gamma^{th}$ moment
of computation, for $\gamma \in [0,1]$ is:
    \begin{equation} \lambda E_0(\gamma) > E_s(\gamma) \label{eqn:jsccondition}\end{equation}
The condition of (\ref{eqn:jsccondition}) has a matching converse once again, which can be
found in \cite{ArikanSourceChannel}.

In section \ref{sec:biasforcomp}, it is shown that the error exponent is positive when the bias
is set as suggested in Theorems \ref{thm:compsi} and \ref{thm:compsc}. Hence, the decoder is
actually decoding correctly and the average computation is not finite simply because the Stack
Algorithm is blindly following an incorrect path.

\subsection{Proof Outline} \label{sec:proofoutline}
    The proofs are the source coding analog of the proofs of
Theorems 2 and 3 of \cite{JelinekUpperBound}. We give a proof outline for Theorem
\ref{thm:proberrorsi} for the point-to-point lossless source coding case, as the important
ideas are all present without the excess notation.

Assume that $G \leq \frac{1+\rho}{\rho}[E_s(\rho) - F_s(\rho)]$. We will show that for any
$\eps > 0$, there is a $K_\eps < \infty$,
\begin{equation} P_e(d) < K_\eps \exp_2 \b( -d (\rho R - E_s(\rho) - \eps ) \b) \end{equation}
We can assume $\rho R - E_s(\rho) - \eps > 0$; otherwise there is nothing to prove.

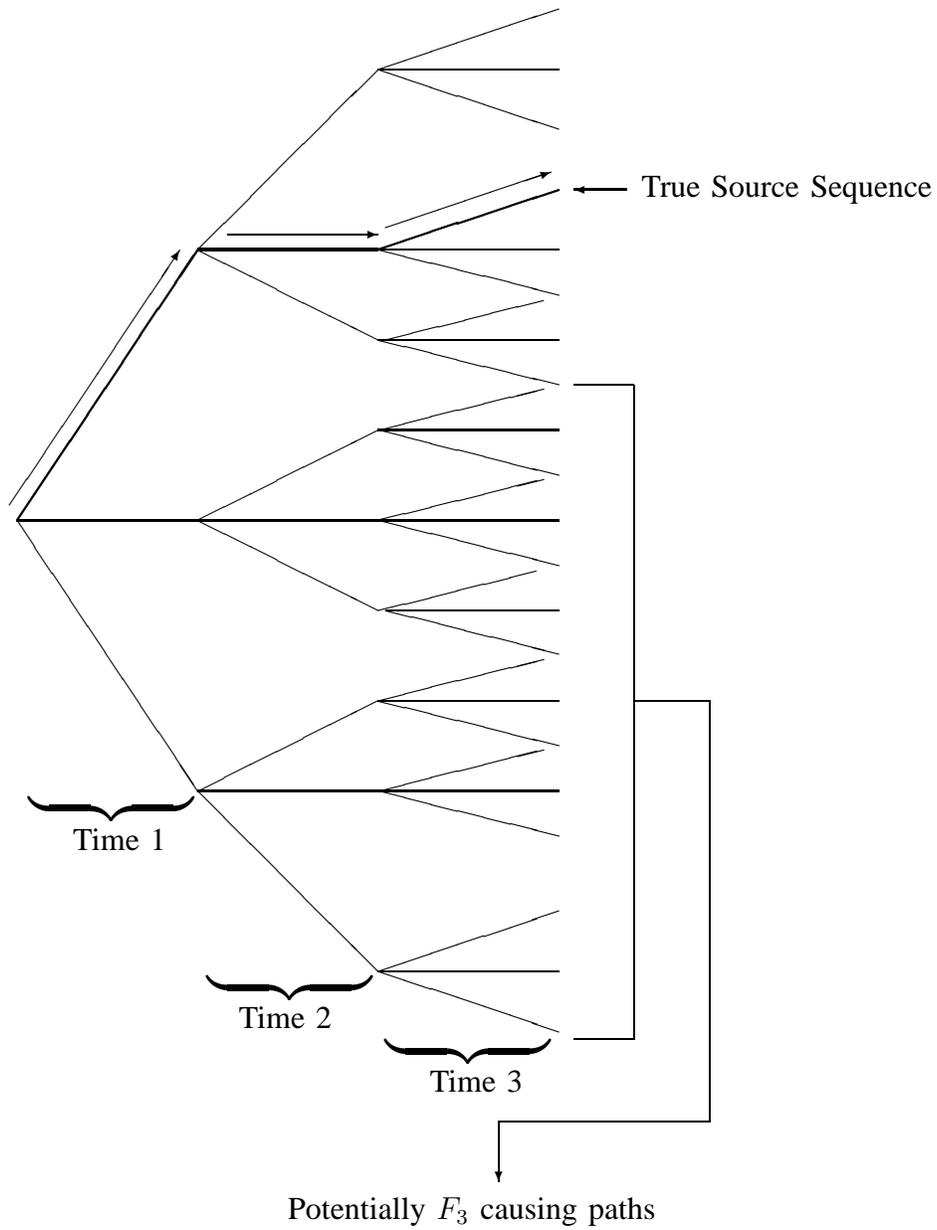
\begin{figure}[htbp]
\begin{center}

\setlength{\unitlength}{3158sp}%
\begingroup\makeatletter\ifx\SetFigFont\undefined%
\gdef\SetFigFont#1#2#3#4#5{%
  \reset@font\fontsize{#1}{#2pt}%
  \fontfamily{#3}\fontseries{#4}\fontshape{#5}%
  \selectfont}%
\fi\endgroup%
\begin{picture}(7161,9612)(629,-8987)
\thinlines \put(651,-3295){\vector( 2, 3){1338}} \put(2363,-1169){\vector( 1, 0){1181}}
\put(3603,-1110){\vector( 3, 1){1299}} \put(710,-3413){\line( 2, 3){1418}}
\put(710,-3413){\line( 1, 0){1417}} \put(710,-3413){\line( 2,-3){1418}} \put(2127,-1287){\line(
1, 0){1417}} \put(2127,-1287){\line( 1, 1){1417}} \put(2127,-1287){\line( 2,-1){1418}}
\put(2127,-3413){\line( 1, 0){1417}} \put(2127,-3413){\line( 2, 1){1418}}
\put(2127,-3413){\line( 2,-1){1418}} \put(2127,-5539){\line( 1, 0){1417}}
\put(2127,-5539){\line( 2, 1){1418}} \put(2127,-5539){\line( 1,-1){1417}} \put(3544,130){\line(
1, 0){1418}} \put(3544,130){\line( 3, 1){1419}} \put(3544,130){\line( 3,-1){1419}}
\put(3544,-1287){\line( 3, 1){1419}}
\put(3544,-1287){\makebox(2.0833,14.5833){\SetFigFont{5}{6}{\rmdefault}{\mddefault}{\updefault}.}}
\put(3544,-1287){\line( 1, 0){1418}} \put(4962,-1287){\line(-1, 0){ 59}}
\put(3544,-1287){\line( 4,-1){1420}} \put(3544,-1996){\line( 1, 0){1418}}
\put(3544,-1996){\line( 4,-1){1420}}
\put(3603,-1996){\makebox(2.0833,14.5833){\SetFigFont{5}{6}{\rmdefault}{\mddefault}{\updefault}.}}
\put(3603,-1996){\line( 4, 1){1240}}
\put(3544,-2704){\makebox(2.0833,14.5833){\SetFigFont{5}{6}{\rmdefault}{\mddefault}{\updefault}.}}
\put(3544,-2704){\line( 1, 0){1418}} \put(3544,-2704){\line( 4, 1){1300}}
\put(3544,-2704){\line( 4,-1){1420}} \put(3544,-3413){\line( 1, 0){1418}}
\put(3544,-3413){\line( 4, 1){1300}} \put(3544,-3413){\line( 4,-1){1420}}
\put(3544,-4122){\line( 4, 1){1240}} \put(3603,-4122){\line( 4,-1){1360}}
\put(3603,-4122){\line( 1, 0){1359}} \put(3544,-4830){\line( 1, 0){1418}}
\put(3544,-4830){\line( 4, 1){1300}} \put(3544,-4830){\line( 4,-1){1420}}
\put(3544,-5539){\line( 1, 0){1418}} \put(3544,-5539){\line( 4, 1){1300}}
\put(3544,-5539){\line( 4,-1){1420}} \put(3544,-6956){\line( 1, 0){1418}}
\put(3544,-6956){\line( 3, 1){1419}} \put(3544,-6956){\line( 3,-1){1419}}
\put(5080,-2350){\line( 1, 0){472}} \put(5552,-2350){\line( 0,-1){5138}}
\put(5552,-7488){\line(-1, 0){354}} \put(5198,-7488){\line(-1, 0){118}}
\put(5493,-815){\vector(-1, 0){413}} \thicklines \put(710,-3413){\line( 2, 3){1418}}
\put(2128,-1286){\line( 1, 0){1416}} \put(3544,-1286){\line( 3, 1){1419}} \thinlines
\put(5552,-4830){\line( 1, 0){591}} \put(6143,-4830){\line( 0,-1){3307}}
\put(6143,-8137){\line(-1, 0){1654}} \put(4489,-8137){\vector( 0,-1){473}}
\put(800,-5590){\makebox(0,0)[lb]{\smash{{\SetFigFont{16}{13.2}{\rmdefault}{\mddefault}{\updefault}$\underbrace{{\phantom{aaaaaaa}}}$}}}}
\put(2200,-7000){\makebox(0,0)[lb]{\smash{{\SetFigFont{16}{13.2}{\rmdefault}{\mddefault}{\updefault}$\underbrace{{\phantom{aaaaaaa}}}$}}}}
\put(3600,-7503){\makebox(0,0)[lb]{\smash{{\SetFigFont{16}{13.2}{\rmdefault}{\mddefault}{\updefault}$\underbrace{{\phantom{aaaaaaa}}}$}}}}
\put(2836,-8905){\makebox(0,0)[lb]{\smash{{\SetFigFont{12}{16.8}{\rmdefault}{\mddefault}{\updefault}Potentially
$F_3$ causing paths}}}}
\put(5611,-874){\makebox(0,0)[lb]{\smash{{\SetFigFont{12}{13.2}{\rmdefault}{\mddefault}{\updefault}True
Source Sequence}}}}
\put(1150,-6000){\makebox(0,0)[lb]{\smash{{\SetFigFont{12}{13.2}{\rmdefault}{\mddefault}{\updefault}Time
1}}}}

\put(2450,-7400){\makebox(0,0)[lb]{\smash{{\SetFigFont{12}{13.2}{\rmdefault}{\mddefault}{\updefault}Time
2}}}}

\put(3950,-7900){\makebox(0,0)[lb]{\smash{{\SetFigFont{12}{13.2}{\rmdefault}{\mddefault}{\updefault}Time
3}}}}
\end{picture}%
\end{center}
\label{fig:sourceerrorevent} \caption{A ternary tree. The true source sequence is shown above.
The condition $\wt{u}_1 \neq u_1$ selects a portion of the tree containing paths that could
potentially cause the error event $F_3$.}
\end{figure}

The error event of interest, $F_d$, is referred to in \cite{JelinekUpperBound} as a failure
event of depth $d$ and is defined in (\ref{eqn:simpleerrorevent}) and we will relate it to
$P_e(d)$ at the end of the proof. Figure \ref{fig:sourceerrorevent} shows paths that may lead
to an error event of depth $3$ occurring, i.e. $F_3$.
    \be \label{eqn:simpleerrorevent}
        F_d \triangleq \b \{ \exists \ \wt{u}_1^d \ , \ \wt{u}_1 \neq u_1\
            \b| \ \Gamma(\wt{u}_1^d) \geq \min_{1 \leq k \leq d} \Gamma(u_1^k) \textrm{ and parities of
            $\wt{u}_1^d$ match $B_1^{dR}$} \b \} \ee

The event $F_d$ can be subdivided into events $F_{d,k}$ so that $F_d = \bigcup_{k=1}^{d}
F_{d,k}$, where \be F_{d,k} \triangleq \b \{ \exists \ \wt{u}_1^{d} \ , \wt{u}_1 \neq u_1 \b |
\quad \Gamma(\wt{u}_1^{d}) \geq \Gamma(u_1^k) \textrm{ and parities of $\wt{u}_1^{d}$ match
$B_1^{dR}$}\b \} \ee

Let $U_1^d$ be the random vector of the first $d$ source symbols and let $\wt{u}_1^d$ be an
arbitrary vector in $\mU^d$. By conditioning on the source sequence and applying the union
bound, we get
    \ba
        P(F_d) & = & \sum_{u_1^d \in \mX^{d}} Q(u_1^{d}) P(F_d|U_1^{d} = u_1^{d})\\
        & \leq & \sum_{k=1}^{d} \sum_{u_1^{d}} Q(u_1^{d})  P(F_{d,k}|U_1^{d} = u_1^{d})
    \ea

Suppose $\wt{u}_1^{d}$ is a false path that causes $F_{d,k}$ to occur. This means its parities
match the received bits and its metric $\Gamma(\wt{u}_1^{d})$ is at least $\Gamma(u_1^k)$.
Therefore,
    \ba
        0 & \leq & \Gamma(\wt{u}_1^{d}) - \Gamma(u_1^k) \\
        & = & \sum_{l=1}^k \log_2\b(\frac{Q(\wt{u}_l)}{Q(u_l)}\b) + \sum_{l = k+1}^{d} \log_2
            Q(\wt{u}_l) + (d - k)G
    \ea

Now, denoting $1(\cdot)$ as the indicator function of its argument, and using a Gallager-style
union bound, for $\rho \in [0,1]$, we have
    \ba
        P(F_{d,k}|U_1^d = u_1^d) & \leq & E \B[ \b (\sum_{\wt{u}_1^d:\ \wt{u}_1 \neq u_1}
        1(\textrm{$\wt{u}_1^d$ causes $F_{d,k}$ to occur}) \b )^\rho \B | U_1^d = u_1^d\B] \\
        & \leq & \B( \sum_{\wt{u}_1^d, \ \wt{u}_1 \neq u_1} E\b[1(\textrm{$\wt{u}_1^d$ causes $F_{d,k}$ to occur})\b|U_1^d = u_1^d\b] \B)^\rho \label{eqn:concavity} \\
        & \triangleq & \B( \sum_{\wt{u}_1^d, \ \wt{u}_1 \neq u_1} A_k(\wt{u}_1^d, u_1^d) \B)^\rho
    \ea
Here (\ref{eqn:concavity}) follows from Jensen's inequality. Continuing with the bounding, we
use the fact that the parity generation process is independent\footnote{Note that we need only
pairwise independence of the parities along two paths.} of everything else to get
    \ba
        A_k(\wt{u}_1^d, u_1^d) & =& E \b[ 1(\textrm{parities of $\wt{u}_1^d$ match $B_1^{dR}$}) \cdot 1(\Gamma(\wt{u}_1^d) \geq \Gamma(u_1^k)) \b| U_1^d =
        u_1^d \b] \\
        & =& E\b[1(\textrm{parities of $\wt{u}_1^d$ match $B_1^{dR}$})\b]E\b[ 1(\Gamma(\wt{u}_1^d) \geq \Gamma(u_1^k)) \b| U_1^d =
        u_1^d \b] \\
        & \leq & \exp_2(-dR)\cdot \exp_2\b(\frac{1}{1+\rho}( \Gamma(\wt{u}_1^d) - \Gamma(u_1^k)) \b)
    \ea

Substituting for $A_k(\wt{u}_1^d,u_1^d)$, and removing the restriction that $\wt{u}_1 \neq
u_1$,
    \ba
        P(F_{d,k}|u_1^d) & \leq & \B( \sum_{\wt{u}_1^d} \exp_2(-dR) \exp_2\b(\frac{1}{1+\rho} \b[
            \log_2\frac{Q(\wt{u}_1^k)}{Q(u_1^k)} + \log_2 Q(\wt{u}_{k+1}^d) + (d-k)G\b] \b)
            \B)^\rho  \\
        & = &  \exp_2\b(-d\rho R + (d-k)\frac{\rho}{1+\rho} G\b) \B(\sum_{\wt{u}_1^k} \b (\frac{Q(\wt{u}_1^k)}{Q(u_1^k)}
            \b )^\frac{1}{1+\rho} \B )^\rho \B(\sum_{\wt{u}_{k+1}^d} Q(\wt{u}_{k+1}^d)^\frac{1}{1+\rho} \B )^\rho \label{eqn:stdalgebra}
    \ea
Equation (\ref{eqn:stdalgebra}) follows from the standard algebra of interchanging sums and
products. Finally, we are ready to complete the bound of $P(F_d)$.
    \ba
        P(F_d) & \leq & \sum_{k=1}^d 2^{-d\rho R + (d-k)\frac{\rho}{1+\rho} G} \nonumber \\
            & & \hspace{15mm} \sum_{u_1^k} Q(u_1^k)^\frac{1}{1+\rho} \b(\sum_{\wt{u}_1^k} Q(\wt{u}_1^k)^\frac{1}{1+\rho}
            \b)^\rho \sum_{u_{k+1}^d} Q(u_{k+1}^d) \b (\sum_{\wt{u}_{k+1}^d} Q(\wt{u}_{k+1}^d)^\frac{1}{1+\rho} \b)^\rho \\
        & = & \sum_{k=1}^d 2^{-d\rho R + (d-k)\frac{\rho}{1+\rho} G} \b(\sum_{u_1^k} Q(u_1^k)^\frac{1}{1+\rho}\b)
            \b(\sum_{u_1^k} Q(u_1^k)^\frac{1}{1+\rho} \b)^\rho \b (\sum_{u_{k+1}^d} Q(u_{k+1}^d)^\frac{1}{1+\rho} \b)^\rho
        \label{eqn:dummyvariable}
    \ea
We get (\ref{eqn:dummyvariable}) by noting that the $\wt{u}$'s are just dummy variables and we
are free to replace them with $u$'s. Next, we use the IID property of the source along with
standard algebra to get to an exponential form. For example, we have
    \ba
       \B(\sum_{u_1^k} Q(u_1^k)^\frac{1}{1+\rho}\B)^{1+\rho} & = & \B( \sum_{u\in \mU} Q(u)^\frac{1}{1+\rho} \B)^{k(1+\rho)} \\
            & = & \exp_2\b( k E_s(\rho) \b) \ea

Similarly,
    \ba
        \b( \sum_{u_1^k} Q(u_1^k)^\frac{1}{1+\rho} \b)^\rho & = & \b( \sum_{u \in \mU} Q(u)^\frac{1}{1+\rho} \b)^{k\rho} \\
        & = & \exp_2 \b( k F_s(\rho) \b) \ea

A bit more algebra and the condition on the bias gives:
    \ba
        P(F_d) & \leq & \sum_{k=1}^d \exp_2\b( - d\rho R + (d-k)\frac{\rho}{1+\rho} G + k E_s(\rho) + (d-k) F_s(\rho) \b)  \\
            & = & \exp_2\b(d\big ( \frac{\rho}{1+\rho} G + F_s(\rho) - \rho R \big) \b) \sum_{k=1}^d \exp_2 \b( k \big(
                E_s(\rho) - F_s(\rho) - \frac{\rho}{1+\rho} G \big) \b) \\
            & \leq & \exp_2\b(d\big ( \frac{\rho}{1+\rho} G + F_s(\rho) - \rho R \big) \b)\cdot d \exp_2 \b( d \big( E_s(\rho) - F_s(\rho) - \frac{\rho}{1+\rho} G \big) \b) \\
            & = & d \exp_2 \b( -d(\rho R - E_s(\rho) ) \b) \ea

So, now we have for any $\epsilon > 0$,
    \ba
        P(F_d) & \leq & \wt{K}_\eps \exp_2 \b( -d(\rho R - E_s(\rho) -\epsilon )\b) \\
        \wt{K}_\eps & \triangleq & \max \b \brl d : \frac{\log_2 d}{d} \geq \eps \b \brr < \infty \ea
Note that $\wt{K}_\eps < \infty$ and is independent of $d$ because $\log_2 (d) / d$ goes to
$0$. Finally, we can prove the statement of the theorem. In order for a delay $d$ or greater
error to occur it must be that $\wh{u}_i(n+d) \neq u_i$ for some $1 \leq i \leq n$.  Now,
assuming the bias satisfies the required condition, we have
    \ba P(\wh{u}_1^n (n+d) \neq u_1^n) & = & \sum_{k=0}^{n-1} P(\wh{u}_1^k(n+d) = u_1^k, \ \wh{u}_{k+1} \neq u_{k+1}) \\
        & \leq & \sum_{k =0}^{n-1} P(F_{d+n-k})P(\wh{u}_1^{k}(n+d) = u_1^{k}) \label{eqn:sourcecriticalstep}\\
        & \leq & \sum_{k = 0}^\infty P(F_{d+k}) \\
        & \leq & \sum_{k=0}^\infty \wt{K}_\eps 2^{-(d+k)(\rho R - E_s(\rho) - \eps)} \\
        & = & 2^{-d(\rho R - E_s(\rho) - \eps)} \sum_{k = 0}^\infty \wt{K}_\eps 2^{-k(\rho R - E_s(\rho) - \eps)} \\
        & = & K_\eps \exp_2 \b( -d(\rho R - E_s(\rho) - \eps) \b) \ea

The critical step is in (\ref{eqn:sourcecriticalstep}), which says that if the decoded path and
true path agree until time $k$, the error event can be thought of as `rooted' at time $k+1$.
Hence, we are reduced to the error event $F_{d+n-k}$. The ideas used in the proof of the
computation bound are essentially the same.

\clearpage

\section{Simulations}

    The random time-varying encoder and stack decoder were simulated in software using a
random number generator. The `experimental' results are compared with the theory for
verification.
    The probability of error with delay, $P_e(d)$, is the first quantity looked at experimentally. Since probability of error decays exponentially with delay, the {\em
logarithm} of the probability of error decays linearly with delay. That is,
    \begin{displaymath} \log_2 P_e(d) \sim -E(R)d \end{displaymath}

The slope of the line on a $log_2$-plot is thus the negative of the error exponent achieved by
this scheme.

Further, if we assume that the moments of computation {\em at any time} are the same as the
moments of computation {\em in any incorrect subtree}, we can compare the Pareto exponent of
the simulation to the theory. This is done by comparing $\log_2 P(C \geq n) $ versus $\log_2 n$
on a graph, where $C$ is the number of computations performed at a time step. The fact that the
distribution of computation is asymptotically Paretian should yield that
    \begin{displaymath} \log_2 P(C \geq n) \sim - \gamma \log_2 n \end{displaymath}
where $\gamma$ is the Pareto exponent of computation.

\subsection{Point to point}
\begin{example}\label{ex:binarysource}
    We explore an example of point-to-point lossless source coding that will be comparable to the case when side
information is available at the decoder only. The source $U_i$ is a sequence of IID $\mB(1/2)$
random bits. $V_i$ are generated by passing $U_i$ through a binary symmetric channel (BSC) with
crossover probability $\eps = \frac{1}{10}$. In this example, we consider the case when the
side information is available at both the encoder and decoder. The situation is diagrammed in
Figure \ref{fig:binaryppsourceexample}. It is clear that since $V$ is available at both the
encoder and decoder, compressing $U \oplus V$ is the same as compressing $U$. Figure
\ref{fig:binarysourceexponentcurves} shows the relevant source coding functions for the error
random variable $U \oplus V$. Since we are just encoding the noise, the rate must be at least
$H(U|V) = H_b(\eps)$ where $H_b$ is the binary entropy function.

\begin{figure}[htbp]
\begin{center}
\includegraphics[width=3in]{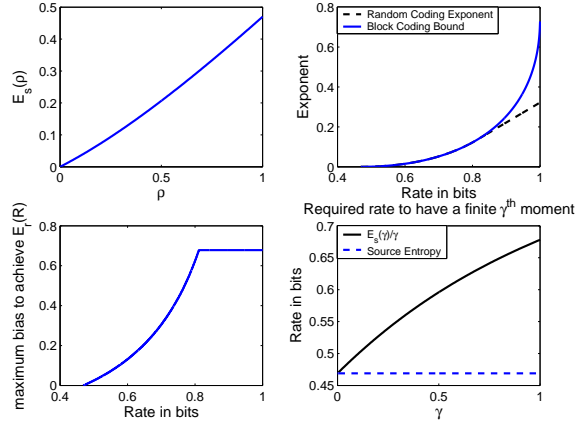}
\end{center}
\caption{Functions of interest associated with a binary source with PMF $(0.9, 0.1)$.}
\label{fig:binarysourceexponentcurves}
\end{figure}

We experimentally estimate the error exponent with delay and Pareto exponent of computation.
These are shown in Figures \ref{fig:binarysourceerrorfit} and \ref{fig:binarysourcecompfit}
respectively. Again, we see that we can achieve the random coding error exponent and the Pareto
exponent guaranteed by theorem \ref{thm:compsi} holds. Since the bias value $(0.7)$ is actually
too high to guarantee achieving $E_{r,pp}(R)$ at rate $R=0.7$, the error exponent in the
experiment is somewhat surprising. However, we stress again that the fitting of a line to the
curve is somewhat arbitrary and we cannot expect to have precise values of the slope beyond the
first digit.

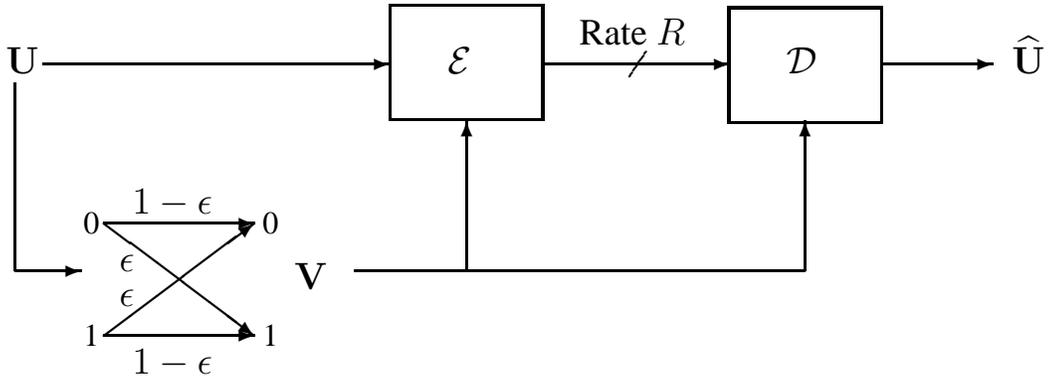
\begin{figure}
\begin{center}

\setlength{\unitlength}{3947sp}%
\begingroup\makeatletter\ifx\SetFigFont\undefined%
\gdef\SetFigFont#1#2#3#4#5{%
  \reset@font\fontsize{#1}{#2pt}%
  \fontfamily{#3}\fontseries{#4}\fontshape{#5}%
  \selectfont}%
\fi\endgroup%
\begin{picture}(6319,2401)(887,-2712)
\thicklines \put(1503,-1696){\vector( 1, 0){945}} \put(1503,-2405){\vector( 1, 0){945}}
\put(1503,-2405){\vector( 4, 3){944}} \put(1503,-1696){\vector( 4,-3){944}}
\put(1680,-1637){\makebox(0,0)[lb]{\smash{{\SetFigFont{14}{16.8}{\rmdefault}{\mddefault}{\updefault}$1-\epsilon$}}}}
\put(1600,-2000){\makebox(0,0)[lb]{\smash{{\SetFigFont{14}{16.8}{\rmdefault}{\mddefault}{\updefault}$\epsilon$}}}}
\put(1600,-2220){\makebox(0,0)[lb]{\smash{{\SetFigFont{14}{16.8}{\rmdefault}{\mddefault}{\updefault}$\epsilon$}}}}
\put(1680,-2641){\makebox(0,0)[lb]{\smash{{\SetFigFont{14}{16.8}{\rmdefault}{\mddefault}{\updefault}$1-\epsilon$}}}}
\put(5434,-1059){\framebox(945,709){}} \put(3308,-1042){\framebox(945,709){}}
\put(946,-815){\line( 0,-1){1181}} \put(946,-1996){\vector( 1, 0){413}} \put(3072,-1996){\line(
1, 0){709}} \put(3781,-1996){\vector( 0, 1){945}} \put(3781,-1996){\line( 1, 0){2126}}
\put(5907,-1996){\vector( 0, 1){945}} \put(1118,-696){\vector( 1, 0){2180}}
\put(4253,-696){\vector( 1, 0){1181}} \put(6379,-696){\vector( 1, 0){709}}
\multiput(4903,-625)(-4.72000,-7.08000){23}{\makebox(6.6667,10.0000){\SetFigFont{7}{8.4}{\rmdefault}{\mddefault}{\updefault}.}}
\put(3662,-755){\makebox(0,0)[lb]{\smash{{\SetFigFont{14}{16.8}{\rmdefault}{\mddefault}{\updefault}$\mathcal{E}$}}}}
\put(5788,-755){\makebox(0,0)[lb]{\smash{{\SetFigFont{14}{16.8}{\rmdefault}{\mddefault}{\updefault}$\mathcal{D}$}}}}
\put(2700,-2114){\makebox(0,0)[lb]{\smash{{\SetFigFont{14}{16.8}{\rmdefault}{\mddefault}{\updefault}$\mathbf{V}$}}}}
\put(4489,-578){\makebox(0,0)[lb]{\smash{{\SetFigFont{14}{16.8}{\rmdefault}{\mddefault}{\updefault}Rate
$R$}}}}
\put(7206,-755){\makebox(0,0)[lb]{\smash{{\SetFigFont{14}{16.8}{\rmdefault}{\mddefault}{\updefault}$\widehat{\mathbf{U}}$}}}}
\put(1375,-1759){\makebox(0,0)[lb]{\smash{{\SetFigFont{12}{14.4}{\rmdefault}{\mddefault}{\updefault}0}}}}
\put(2500,-2468){\makebox(0,0)[lb]{\smash{{\SetFigFont{12}{14.4}{\rmdefault}{\mddefault}{\updefault}1}}}}
\put(2500,-1759){\makebox(0,0)[lb]{\smash{{\SetFigFont{12}{14.4}{\rmdefault}{\mddefault}{\updefault}0}}}}
\put(1375,-2468){\makebox(0,0)[lb]{\smash{{\SetFigFont{12}{14.4}{\rmdefault}{\mddefault}{\updefault}1}}}}
\put(887,-755){\makebox(0,0)[lb]{\smash{{\SetFigFont{14}{16.8}{\rmdefault}{\mddefault}{\updefault}$\mathbf{U}$}}}}
\end{picture}%
\end{center}
\caption[An example of point-to-point source coding that can be compared to source coding with
side information at the decoder]{An example of point-to-point source coding that can be
compared to source coding with side information at the decoder. $U_i$ are Bernoulli (1/2)
random bits, $V$ is $U$ passed through a BSC with crossover probability $\eps$. The encoder
sequentially bins the error sequence $U \oplus V$.} \label{fig:binaryppsourceexample}
\end{figure}

\begin{table*}
\begin{center}
\begin{tabular}{|c|c|c|}
  \hline

  & Theoretical & Experimental \\
  \hline
Error exponent with delay & $0.05$ & $\sim 0.06$\\
Pareto exponent of computation & $\geq 1$  &$\sim 1.2$\\
Conjectured Pareto exponent & 1.2 & \\
  \hline
\end{tabular}

  \caption{Comparison of theoretical and experimental performance in example \ref{ex:binarysource}.}

  \label{table:binarysource}
  \end{center}
\end{table*}

\begin{figure}[!htbp]
\begin{center}
\includegraphics[width=3in]{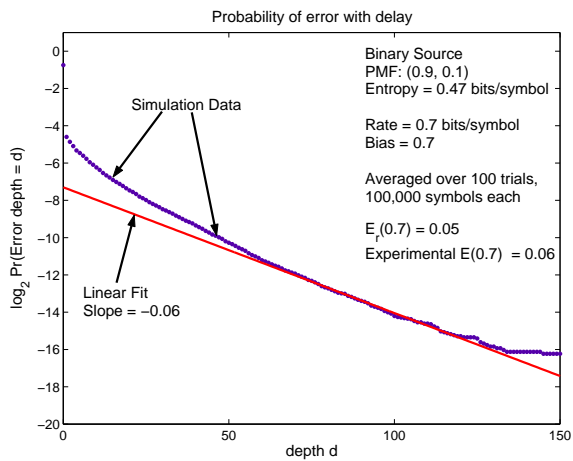}
\end{center}
\caption{Estimating $E(R)$ for example \ref{ex:binarysource}.}
\label{fig:binarysourceerrorfit}\end{figure}

\begin{figure}
\begin{center}
\includegraphics[width=3in]{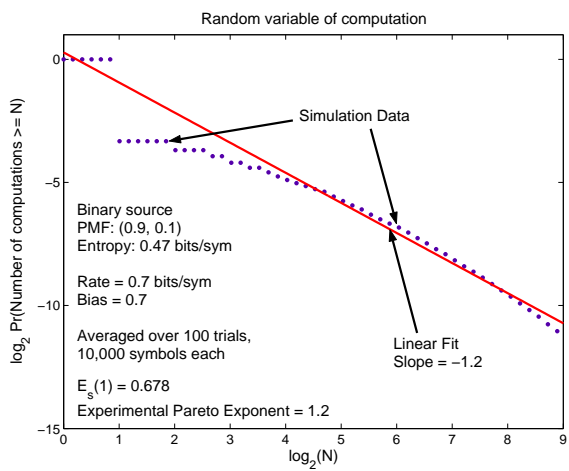}
\end{center}
\caption{Estimating the Pareto exponent for computation for example
\ref{ex:binarysource}.}\label{fig:binarysourcecompfit}
\end{figure}

\end{example}

\subsection{Side information}

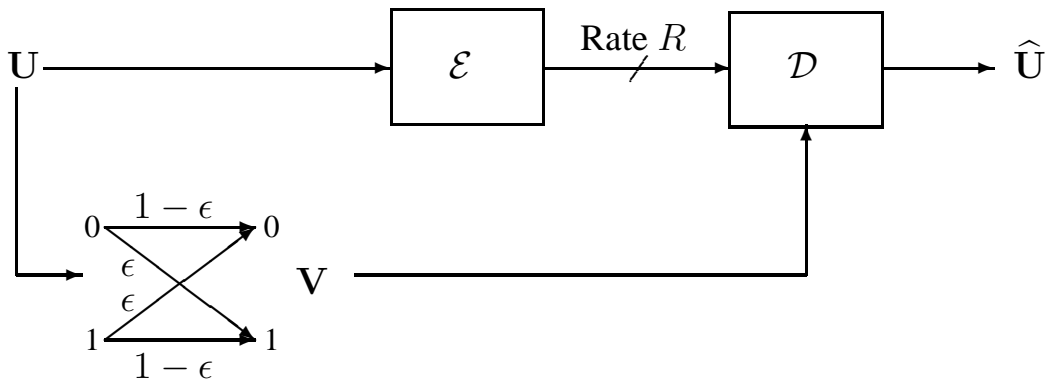
\begin{figure}
\begin{center}

\setlength{\unitlength}{3947sp}%
\begingroup\makeatletter\ifx\SetFigFont\undefined%
\gdef\SetFigFont#1#2#3#4#5{%
  \reset@font\fontsize{#1}{#2pt}%
  \fontfamily{#3}\fontseries{#4}\fontshape{#5}%
  \selectfont}%
\fi\endgroup%
\begin{picture}(6319,2401)(887,-2712)
\thicklines \put(1503,-1696){\vector( 1, 0){945}} \put(1503,-2405){\vector( 1, 0){945}}
\put(1503,-2405){\vector( 4, 3){944}} \put(1503,-1696){\vector( 4,-3){944}}
\put(1680,-1637){\makebox(0,0)[lb]{\smash{{\SetFigFont{14}{16.8}{\rmdefault}{\mddefault}{\updefault}$1-\epsilon$}}}}
\put(1600,-2000){\makebox(0,0)[lb]{\smash{{\SetFigFont{14}{16.8}{\rmdefault}{\mddefault}{\updefault}$\epsilon$}}}}
\put(1600,-2220){\makebox(0,0)[lb]{\smash{{\SetFigFont{14}{16.8}{\rmdefault}{\mddefault}{\updefault}$\epsilon$}}}}
\put(1680,-2641){\makebox(0,0)[lb]{\smash{{\SetFigFont{14}{16.8}{\rmdefault}{\mddefault}{\updefault}$1-\epsilon$}}}}
\put(5434,-1059){\framebox(945,709){}} \put(3308,-1042){\framebox(945,709){}}
\put(946,-815){\line( 0,-1){1181}}
\put(946,-1996){\vector( 1, 0){413}} \put(3072,-1996){\line( 1, 0){709}} 
\put(3781,-1996){\line( 1, 0){2126}} \put(5907,-1996){\vector( 0, 1){945}}
\put(1118,-696){\vector( 1, 0){2180}} \put(4253,-696){\vector( 1, 0){1181}}
\put(6379,-696){\vector( 1, 0){709}}
\multiput(4903,-625)(-4.72000,-7.08000){23}{\makebox(6.6667,10.0000){\SetFigFont{7}{8.4}{\rmdefault}{\mddefault}{\updefault}.}}
\put(3662,-755){\makebox(0,0)[lb]{\smash{{\SetFigFont{14}{16.8}{\rmdefault}{\mddefault}{\updefault}$\mathcal{E}$}}}}
\put(5788,-755){\makebox(0,0)[lb]{\smash{{\SetFigFont{14}{16.8}{\rmdefault}{\mddefault}{\updefault}$\mathcal{D}$}}}}
\put(2700,-2114){\makebox(0,0)[lb]{\smash{{\SetFigFont{14}{16.8}{\rmdefault}{\mddefault}{\updefault}$\mathbf{V}$}}}}
\put(4489,-578){\makebox(0,0)[lb]{\smash{{\SetFigFont{14}{16.8}{\rmdefault}{\mddefault}{\updefault}Rate
$R$}}}}
\put(7206,-755){\makebox(0,0)[lb]{\smash{{\SetFigFont{14}{16.8}{\rmdefault}{\mddefault}{\updefault}$\widehat{\mathbf{U}}$}}}}
\put(1375,-1759){\makebox(0,0)[lb]{\smash{{\SetFigFont{12}{14.4}{\rmdefault}{\mddefault}{\updefault}0}}}}
\put(2500,-2468){\makebox(0,0)[lb]{\smash{{\SetFigFont{12}{14.4}{\rmdefault}{\mddefault}{\updefault}1}}}}
\put(2500,-1759){\makebox(0,0)[lb]{\smash{{\SetFigFont{12}{14.4}{\rmdefault}{\mddefault}{\updefault}0}}}}
\put(1375,-2468){\makebox(0,0)[lb]{\smash{{\SetFigFont{12}{14.4}{\rmdefault}{\mddefault}{\updefault}1}}}}
\put(887,-755){\makebox(0,0)[lb]{\smash{{\SetFigFont{14}{16.8}{\rmdefault}{\mddefault}{\updefault}$\mathbf{U}$}}}}
\end{picture}%
\end{center}
\caption[An example of lossless source coding with side information]{An example of lossless
source coding with side information at the decoder only. $U_i$ are Bernoulli (1/2) random bits,
$V$ is $U$ passed through a BSC with crossover probability $\eps$. The encoder sequentially
bins its observations of $U$.} \label{fig:binarysisourceexample}
\end{figure}

\begin{example}\label{ex:binarysourcesiexample} We reuse the binary source example, where the
side information is generated by passing the source bit through a BSC. The side information
this time is only available at the decoder, as is shown in Figure
\ref{fig:binarysisourceexample}. In this case, the function $E_{si}(\rho)$
simplifies\footnote{This expansion is for the reviewer's convenience, it will be removed in the
final version.} as below,
\begin{eqnarray}
E_{si}(\rho) & = & \log_2 \sum_{v\in \mV} \b ( \sum_{u \in \mU} Q(u,v)^{1/(1+\rho)} \b)^{1+\rho} \b) \\
    & = & \log_2 \sum_{v\in \mV} Q(v) \b ( \sum_{u \in \mU} Q(u|v)^{1/(1+\rho)} \b)^{1+\rho} \b) \\
    & = & \log_2 \sum_{v=0}^1 \frac{1}{2} \b( \eps^{1/(1+\rho)} + (1-\eps)^{1/(1+\rho)} \b )^{1+\rho}\\
    & = & (1+\rho) \log_2 \b ( \eps^{1/(1+\rho)} + (1-\eps)^{1/(1+\rho)} \b )
    \end{eqnarray}

This $E_{si}(\rho)$ is the same as the $E_s(\rho)$ function that appears if the side
information $V$ is available at both the encoder and decoder, i.e. point-to-point coding of the
error sequence. To compare to the case when $V$ is available at the encoder as well, we
estimate the error exponent with delay and the Pareto exponent for computation through
simulation in Figures \ref{fig:binarysourcesierrorfit} and \ref{fig:binarysourcesicompfit}
respectively. In this simulation, the rate is once again $0.7$ bits per symbol, and the bias is
$0.7$. We see nearly identical values for the error exponent and Pareto exponent of the two
examples, as we should.

\begin{table*}
\begin{center}
\begin{tabular}{|c|c|c|}
  \hline

  & Theoretical & Experimental \\
  \hline
Error exponent with delay & $0.05$ & $\sim 0.08$\\
Pareto exponent of computation & $\geq 1$  &$\sim 1.2$\\
Conjectured Pareto exponent & 1.2 &\\
  \hline
\end{tabular}
\end{center}
\caption{Comparison of theoretical and experimental performance in example
\ref{ex:binarysourcesiexample}.} \label{table:binarysourcesi}
\end{table*}

\begin{figure} \begin{center}
    \includegraphics[width = 3in]{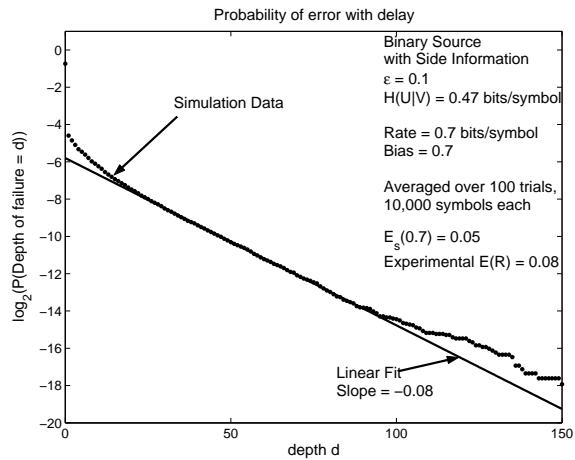}
    \end{center}
    \caption{Determining $E(R)$ for example \ref{ex:binarysourcesiexample}.}
    \label{fig:binarysourcesierrorfit}
\end{figure}
\begin{figure}\begin{center} \includegraphics[width =
    3in]{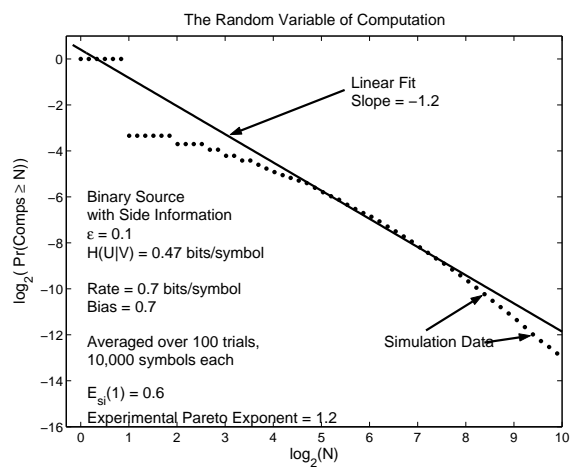}
    \end{center}
    \caption{Determining the Pareto exponent for computation for example
    \ref{ex:binarysourcesiexample}.} \label{fig:binarysourcesicompfit}
\end{figure}

\end{example}

\section{Conclusion}
    In this paper, a scheme was described for the problem of joint source-channel coding with
side information available only at the decoder. If the channel is noiseless, one immediately
arrives at a scheme for (almost) lossless compression with side information at the decoder
only. The coding is done in a `streaming' manner in the sense that source symbols are encoded
as they arrive. The encoder consists of an infinite constraint length random time-varying
convolutional code, and the decoder is a Stack Algorithm sequential decoder with a variable
`bias' parameter.

    Two performance measures were bounded for this system when coding IID sources over DMCs;
probability of error with end-to-end delay and (average) computational effort of the decoder.
We showed that various analogs of Gallager's random coding error exponent could be achieved by
suitable choice of bias. We also bounded the $\rho^{th}$ moment of computation for $0 \leq \rho
\leq 1$. We thus established a lower bound for the cutoff rate for moments up to the mean for
sequential decoding with side information. One would expect that a tweak to the analysis of
\cite{ArikanSourceChannel}, allowing for side information, would establish the matching upper
bounds on the cutoff rate.

    Following the work of Koshelev \cite{Koshelev}, it may be possible to even allow for
finite memory Markov sources. Another important extension would be to consider two distributed
encoders as in the paper of Slepian and Wolf \cite{SlepianWolf}; the case when the side
information $V$ is coded and required to be reconstructed. The scheme of Section
\ref{sec:binningscheme} naturally allows for this by adding another tree code for the other
source and modifying the metric update slightly. Simulation results have shown that the
computation cost seems to be prohibitive except for high rates. Indeed, even the random coding
exponents for correlated sources are generally much lower when both sources are coded
\cite{ChangISIT05}. Perhaps this is not surprising considering that the computational cutoff
rate is closely tied to the `Gallager' $E_s$ function indirectly through the random coding
error exponent.

\clearpage

\section{Appendix - Proofs}

In this section we prove the theorems of the paper. First we show that the probability of error
goes to zero exponentially with delay. This is done initially in the case when there is only a
noiseless channel and the source is encoded at some rate $R$ bits per time unit. Then, we prove
this for joint source-channel coding with side information when the source and the channel are
`synchronized' at one source symbol per channel use. Next, we prove the theorems regarding the
random variable of computation. Again, we do this first in the case of source coding with side
information, and then for joint source-channel coding with side information. Before diving into
the proofs individually, we first examine the error events that show up.\footnote{The appendix
is lengthy and somewhat redundant for the convenience of the reviewer and will be trimmed for
the final version.}

Assume that $R$ is an integer so we need not worry about integer effects \footnote{For a
non-integer rate, the encoder outputs either $\lfloor R \rfloor$ or $\lceil R \rceil$ bits at
every time instant. The integer effect is not important asymptotically, and for convenience we
have used the integer assumption in proofs. In simulations, we have used non-integer rates.} in
the exposition, but the results hold for non-integer rates as well. Similarly, assume in the
proof of Theorems \ref{thm:proberrorsc} and \ref{thm:compsc} that $\lambda$ is an integer.

\begin{figure}
\begin{center}
\setlength{\unitlength}{2960sp}%
\begingroup\makeatletter\ifx\SetFigFont\undefined%
\gdef\SetFigFont#1#2#3#4#5{%
  \reset@font\fontsize{#1}{#2pt}%
  \fontfamily{#3}\fontseries{#4}\fontshape{#5}%
  \selectfont}%
\fi\endgroup%
\begin{picture}(6003,1000)(1500,-2081)
\thicklines \put(3781,-1759){\framebox(1417,708){}} \put(1836,-1759){\framebox(1181,708){}}
\put(1100,-1405){\vector( 1, 0){709}} \put(3042,-1405){\vector( 1, 0){720}}
\put(5200,-1405){\vector( 1, 0){670}} \put(5869,-1759){\framebox(1181,708){}}
\put(7068,-1405){\vector( 1, 0){708}} \put(6450, -2259){\vector(0,1){500}}
\put(1300,-1287){\makebox(0,0)[lb]{\smash{{\SetFigFont{14}{16.8}{\rmdefault}{\mddefault}{\updefault}$U$}}}}
\put(2347,-1523){\makebox(0,0)[lb]{\smash{{\SetFigFont{14}{16.8}{\rmdefault}{\mddefault}{\updefault}$\mathcal{E}$}}}}
\put(3190,-1287){\makebox(0,0)[lb]{\smash{{\SetFigFont{14}{16.8}{\rmdefault}{\mddefault}{\updefault}$X$}}}}
\put(5434,-1287){\makebox(0,0)[lb]{\smash{{\SetFigFont{14}{16.8}{\rmdefault}{\mddefault}{\updefault}$Y$}}}}
\put(6350,-2557){\makebox(0,0)[lb]{\smash{{\SetFigFont{14}{16.8}{\rmdefault}{\mddefault}{\updefault}$V$}}}}
\put(6343,-1523){\makebox(0,0)[lb]{\smash{{\SetFigFont{14}{16.8}{\rmdefault}{\mddefault}{\updefault}$\mathcal{D}$}}}}
\put(4353,-1523){\makebox(0,0)[lb]{\smash{{\SetFigFont{14}{16.8}{\rmdefault}{\mddefault}{\updefault}$W$}}}}
\put(7414,-1274){\makebox(0,0)[lb]{\smash{{\SetFigFont{14}{16.8}{\rmdefault}{\mddefault}{\updefault}$\hat{U}$}}}}
\end{picture}%
\end{center}
\caption{Joint source-channel coding with side information available to the decoder.}
\end{figure}
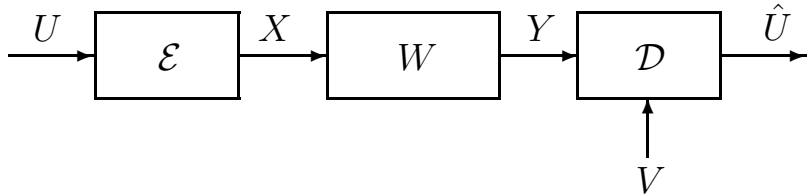

\subsection{Error events}
A source produces IID letters $(U_i,V_i)$ according to a joint distribution $Q(u,v)$ on a
discrete alphabet $\mU \times \mV$. The $U_i$ are available to an encoder, and the $V_i$ are
given to the decoder as side information. In the case of joint source-channel coding, there is
a discrete memoryless channel with probability transition matrix $W(y|x)$ with finite input and
output alphabets. We use the encoder and decoder of Section \ref{sec:binningscheme}. For joint
source-channel coding, we assume there is one channel use for every source symbol. We denote
vectors as $u_1^d, y_n^m,\ldots$ etc. We reserve the letters $u$, $x$, and $y$ for the `true'
variables and $\wt{u}$, $\wt{x}$ for arbitrary `false' variables.

The probability measure $P$ will  refer to all randomness in the source as well as the randomly
generated encoder. When no confusion arises, $Q$ will be applied to multiple symbols like
$u_1^k$ with the meaning that $Q(u_1^k) = \prod_{l=1}^k Q(u_l)$.

The stack algorithm uses a metric, (implicity a function of the side information, tree code and
channel outputs if there are any), of $\Gamma(u) = \log(Q(u|v)W(y|x(u))/P(y)) + G$ for some
bias $G \in \mathbb{R}$, where $P(y) = \sum_{x} \Beta(x)W(y|x)$. If there is no channel,
$\Gamma(u) = \log(Q(u|v))+ G$ if the parities of the sequence match the parities received by
the decoder. Otherwise, we can set the metric for non-matching parities to be $-\infty$ to
effectively drop them out of the stack. We now consider how the stack decoder could follow a
false path. We say the stack decoder `visits' a node if it computes a metric for that node.

Suppose the true source sequence is $(u_1^n,v_1^n)$ until time $n$ and $\wt{u}_1^n$ is some
other arbitrary source sequence. Viewed as paths through the encoding tree, $u_1^n$ and
$\wt{u}_1^n$ are the same if and only if they trace the same path from the root to depth $n$ in
the tree. Also, if they are not the same, there is some earliest point at which they diverge,
call that time $n-d+1$. Equivalently, $u_1^{n-d} = \wt{u}_1^{n-d}$, but $u_1^{n-d+1} \neq
\wt{u}_1^{n-d+1}$. Until time $n-d$, because the stack decoder is a sequential decoder, the
stack algorithm assigns $u_1^{n-d}$ and $\wt{u}_1^{n-d}$ the same metric. In order for
$\wt{u}_1^n$ to be the decoder's estimated path at time $n$, a necessary condition is:
    \begin{equation}
        \Gamma(\wt{u}_1^n) \geq \min_{n-d+1 \leq k \leq n} \Gamma(u_1^k)
    \end{equation}

Noting that $\Gamma(\wt{u}_1^{n-d}) = \Gamma(u_1^{n-d})$, and the fact that the metric is
additive, this reduces to:
    \begin{equation}
        \Gamma(\wt{u}_{n-d+1}^n) \geq \min_{1 \leq k \leq d}
        \Gamma(u_{n-d+1}^{n-d+k})
    \end{equation}

All randomness in the source, encoder/decoder, and channel is memoryless and stationary, so the
probability of the above event occurring for some false $\wt{u}_{n-d+1}^n$ is the same as the
probability of the event $F_d$ defined below:
    \begin{equation}
        F_d  \triangleq  \b\{\exists \wt{u}_1^d \in \mU^d \b | \wt{u}_1 \neq u_1,~
        \Gamma(\wt{u}_1^d) \geq \min_{1 \leq k \leq d} \Gamma(u_1^k) \b\}
    \end{equation}

We call $F_d$ the error event of depth $d$. Figure \ref{fig:sourceerrorevent} shows paths that
may lead to an error event of depth $3$ occurring, i.e. $F_3$. We can further break up $F_d$
into sub-events $F_{d,k}$ so that:
    \begin{eqnarray}
        F_{d,k}& \triangleq & \b\{\exists \wt{u}_1^d \in \mU^d \b | \wt{u}_1 \neq u_1,~
        \Gamma(\wt{u}_1^d) \geq \Gamma(u_1^k) \b\}\\
        P(F_d) & \leq & \sum_{k=1}^d P(F_{d,k}) \\
        P(F_{d,k}) & = & E \b[ 1\b( \exists\wt{u}_1^d \in \mU^d : \wt{u}_1 \neq u_1,~
        \Gamma(\wt{u}_1^d) \geq \Gamma(u_1^k) \b) \b] \\
        & \leq & E\b[ \b( \sum_{\wt{u}_1^d \in \mU^d, \wt{u}_1 \neq u_1} 1(\Gamma(\wt{u}_1^d) \geq
        \Gamma(u_1^k)) \b)^\rho\b] \\
        & & \forall \rho \in [0,1] \nonumber
    \end{eqnarray}

Here $1(\cdot)$ denotes the indicator function of its argument. The last line is in fact true
for any $\rho \geq 0$, but it is only useful in bounding if $\rho \in [0,1]$.

The probability of error with delay $d$ at time $n$ is $P(\wh{u}_1^{n-d}(n) \neq u_1^{n-d})$,
where $\wh{u}_1^{n-d}(n)$ is the decoder's estimate of the source from time $1$ to $n-d$
produced at time $n$. We will give an upper bound on the probability of error independent of
$n$ and depending only on $d$, which is an upper bound on $P_e(d)$.

If $\wh{u}_1^{n-d}(n) \neq u_1^{n-d}$, then there is some point at which they diverged, say
$n-d-l$. So $\wh{u}_1^{n-d-l}(n) = u_1^{n-d-l}$, but $\wh{u}_{n-d-l+1}(n) \neq u_{n-d-l+1}$. So
the probability that a false decoded path and the true path diverged at time $n-d-l$ is at most
$P(F_{d+l})$. Now we can use the union bound to get:
    \begin{equation} P(\wh{u}_1^{n-d}(n) \neq u_1^{n-d}) \leq  \sum_{l=0}^{n-d} P(F_{d+l}) \end{equation}

To get a bound independent of $n$, we just set $n$ to infinity and get
    \begin{equation} P_e(d)  \leq  \sum_{l=0}^\infty P(F_{d+l})\end{equation}

As for the random variable of computation, we define a generic variable $N$ below
\footnote{Sums of the form $\sum_{u_1^l}$ mean summing over all $u_1^l \in \mU^l$. This is the
meaning for all sums in the appendix, unless an additional condition such as $\wt{u}_1 \neq
u_1$ is explicitly stated.}.
    \begin{eqnarray}
        N & \triangleq & \sum_{l=1}^\infty \sum_{\wt{u}_1^l:~ \wt{u}_1 \neq u_1} 1\b(\wt{u}_1^l
        \textrm{ is visited}\b) \\
        & \leq & \sum_{l=1}^\infty \sum_{\wt{u}_1^l:~ \wt{u}_1 \neq u_1} 1\b(\Gamma(\wt{u}_1^l) \geq
        \min_{1 \leq k < \infty} \Gamma(u_1^k) \b)
    \end{eqnarray}

By symmetry, it is clear that $E[N^\gamma] = E[N_i^\gamma]$ for all $i \geq 1$ and any $\gamma
\geq 0$. We want to find when $E[N^\gamma] < \infty$. By concavity, we have
    \begin{eqnarray}
        E[N^\gamma] & \leq & E \h[ \B( \sum_{l=1}^\infty \sum_{\wt{u}_1^l:~ \wt{u}_1 \neq u_1}
        1\b(\Gamma(\wt{u}_1^l) \geq \min_{1 \leq k < \infty} \Gamma(u_1^k)\B) \B)^\gamma\h] \\
        & \leq & E \h[ \B( \sum_{l=1}^\infty \sum_{k=1}^\infty \sum_{\wt{u}_1^l:~ \wt{u}_1 \neq u_1}
        1\b(\Gamma(\wt{u}_1^l) \geq \Gamma(u_1^k) \B) \B)^\gamma\h] \\
        & \leq &  \sum_{l=1}^\infty \sum_{k=1}^\infty  E \h[ \B( \sum_{\wt{u}_1^l:~ \wt{u}_1 \neq u_1}
        1\b(\Gamma(\wt{u}_1^l) \geq \Gamma(u_1^k)\B) \B)^\gamma\h] \\
        & = & \sum_{l=1}^\infty \sum_{k=1}^\infty  A_{l,k} \\
        A_{l,k} & \triangleq & E \h[ \B( \sum_{\wt{u}_1^l:~ \wt{u}_1 \neq u_1} 1\b(\Gamma(\wt{u}_1^l)
        \geq \Gamma(u_1^k) \B)\B)^\gamma\h]
    \end{eqnarray}

Here are some further facts/definitions that are repeatedly used in the appendix:
    \begin{enumerate}
        \item The source and channel are memoryless. The parity generation process and channel input
        generation process are done IID for every branch/node.

        \item Jensen's inequality. If $X$ is a random variable and $f$ is a concave $\cap$
        function, $E[f(X)] \leq f(E[X])$. If $\rho \in [0,1]$, $f(x) = x^\rho$ is concave
        $\cap$.

        \item By definition, for each $y \in \mY$, $P(y) = \sum_{x \in \mX} \Beta(x)W(y|x)$.

        \item Definitions of the exponent functions $E_{si}, E_0$, etc. can be found in
        \ref{sec:gallagerfunctions}.

        \item Sums and products of probabilities commute, and changing dummy variables can be used to simplify terms. See Gallager \cite{GallagerBook},
        Chapter 5.
    \end{enumerate}

\subsection{Probability of error - source coding with side information} \label{sec:proberrorsi}

\begin{theorem}[Restatement of Theorem \ref{thm:proberrorsi}] Suppose that the decoder has access to the side information
and there is a noiseless rate $R$ binary channel between the encoder and decoder. Fix any $\eps
> 0$ and let $\rho \in [0,1]$. For the encoder/decoder of Section \ref{sec:binningscheme}, if the bias $G$ satisfies
\begin{equation} G \leq \frac{1+\rho}{\rho}\B[ E_{si}(\rho) - F_{si}(\rho) \B]
\label{eqn:biascondsi}
\end{equation}
then, there is a constant $K_\eps < \infty$ so that
\begin{equation} P_e(d) < K_\eps \exp_2 \B(-d\b(\rho R - E_{si}(\rho) - \eps\b)\B)
\end{equation}
Hence, with suitable choice of bias, the error exponent with delay can be
\begin{equation} E(R) = E_{r,si}(R) \triangleq \sup_{\rho \in [0,1]} \rho R - E_{si}(\rho)
\end{equation}   \end{theorem}\vspace{.2in}

\begin{proof} The letter $B$ will be used for the bits received by the decoder, which will be referred
to as `parities'. We can specialize the event $F_d$ to this situation and write it as:
    \begin{equation} \label{eqn:deferrorevent}
        F_d \triangleq \b \{ \exists \ \wt{u}_1^d \ , \ \wt{u}_1 \neq u_1\ \b| \
        \Gamma(\wt{u}_1^d) \geq \min_{1 \leq k \leq d} \Gamma(u_1^k) \textrm{ and parities of
        $\wt{u}_1^d$ match $B_1^{dR}$} \b \}
    \end{equation}

The event $F_d$ can be subdivided into events $F_{d,k}$ so that $F_d = \bigcup_{k=1}^{d}
F_{d,k}$, where
    \begin{equation}
        F_{d,k} \triangleq \b \{ \exists \ \wt{u}_1^{d} \ , \wt{u}_1 \neq u_1 \b
        | \quad \Gamma(\wt{u}_1^{d}) \geq \Gamma(u_1^k) \textrm{ and parities of $\wt{u}_1^{d}$ match
        $B_1^{dR}$}\b \}
    \end{equation}

Suppose $ \wt{u}_1^{d}$ is a false path that causes $F_{d,k}$ to occur. This means its parities
match the received bits and its metric $\Gamma( \wt{u}_1^{d})$ is at least $\Gamma(u_1^k)$.
Therefore,
    \begin{eqnarray}
        0 & \leq & \Gamma( \wt{u}_1^{d})
        - \Gamma(u_1^k) \\
        & = & \sum_{l = 1}^{d} \b(\log(Q( \wt{u}_l|v_l)) + G\b) - \sum_{l=1}^k \b(\log(Q(u_l|v_l)) + G \b) \\
        & = & \sum_{l=1}^k \log\b(\frac{Q( \wt{u}_l|v_l)}{Q(u_l|v_l)}\b) + \sum_{l = k+1}^{d} \log Q(
        \wt{u}_l|v_l) + (d - k)G
    \end{eqnarray}

Using a Gallager-style union bound, for $\rho \in [0,1]$, we have
    \begin{eqnarray}
        P(F_{d,k}) & \leq & E \B[ \b ( \sum_{\wt{u}_1^d:\ \wt{u}_1 \neq x_1}
        1(\textrm{$\wt{u}_1^d$ causes $F_{d,k}$ to occur}) \b )^\rho \B]  \label{eqn:Alkscsi}\\
        & \stackrel{(a)}{\leq} & \B( \sum_{\wt{u}_1^d, \ \wt{u}_1 \neq x_1} E\b[1(\textrm{$\wt{u}_1^d$ causes $F_{d,k}$ to occur})\b] \B)^\rho \label{eqn:jensens}
    \end{eqnarray}

Here, (a) is by Jensen's inequality. By conditioning on the source sequence and applying the
union bound, we get
    \begin{eqnarray}
        P(F_d) & = & \sum_{u_1^d
        \in \mU^{d}, v_1^d \in \mV^d} Q(u_1^{d}, v_1^d) P(F_d|u_1^{d}, v_1^d) \\
        & \leq & \sum_{u_1^{d}, v_1^d} Q(u_1^{d}, v_1^d) \sum_{k=1}^{d} P(F_{d,k}|u_1^{d}, v_1^d) \\
        & = & \sum_{k=1}^{d} \sum_{u_1^{d}, v_1^d} Q(u_1^{d}, v_1^d)  P(F_{d,k}|u_1^{d}, v_1^d)
        \\
        & \leq & \sum_{k=1}^d\sum_{u_1^{d}, v_1^d} Q(u_1^{d}, v_1^d) \B( \sum_{\wt{u}_1^d, \ \wt{u}_1 \neq
        x_1}E\b[1(\textrm{$\wt{u}_1^d$ causes $F_{d,k}$ to occur})\b| u_1^d, v_1^d\b] \B)^\rho\\
        & \triangleq & \sum_{k=1}^d\sum_{u_1^d,v_1^d}Q(u_1^d,v_1^d) \B( \sum_{\wt{u}_1^d, \ \wt{u}_1 \neq x_1} A_k(\wt{u}_1^d, u_1^d, v_1^d)
        \B)^\rho \label{eqn:scsiFd}
    \end{eqnarray}

Continuing with the bounding, we use the fact that the parity generation process is
independent\footnote{Only pairwise independence of the parities along two disjoint paths is
required.} of everything else to get
    \begin{eqnarray}
        A_k(\wt{u}_1^d, u_1^d, v_1^d) &
        =& E \b[ 1(\textrm{parities of $\wt{u}_1^d$ match $B_1^{dR}$}) \cdot 1(\Gamma(\wt{u}_1^d) \geq \Gamma(u_1^k)) \b| u_1^d, v_1^d \b] \\
        & =& E\b[1(\textrm{parities of $\wt{u}_1^d$ match $B_1^{dR}$})\b]E\b[ 1(\Gamma(\wt{u}_1^d) \geq \Gamma(u_1^k)) \b| u_1^d, v_1^d\b] \\
        & = & 2^{-dR} E\b[ 1(\Gamma(\wt{u}_1^d) \geq \Gamma(u_1^k)) \b| u_1^d, v_1^d \b] \\
        & \leq & \exp_2(-dR)\cdot \exp_2\b(s( \Gamma(\wt{u}_1^d) - \Gamma(u_1^k)) \b) \\
        & = & \exp_2(-dR) \cdot \exp_2\b( s\b[ \log_2\frac{Q(\wt{u}_1^k|v_1^k)}{Q(u_1^k|v_1^k)} + \log_2 Q(\wt{u}_{k+1}^d|v_{k+1}^d) + (d-k)G \b] \b)\\
        & & \textrm{for any $s \geq 0$} \nonumber
    \end{eqnarray}

Substituting for $A_k(\wt{u}_1^d, u_1^d, v_1^d)$, and removing the restriction that $\wt{u}_1
\neq u_1$,
    \begin{eqnarray}
        P(F_{d,k}|u_1^d, v_1^d) & \leq & \B( \sum_{\wt{u}_1^d} \exp_2(-dR) \exp_2\b(s \b[
        \log_2\frac{Q(\wt{u}_1^k|v_1^k)}{Q(u_1^k|v_1^k)} + \log_2 Q(\wt{u}_{k+1}^d|v_{k+1}^d) +
        (d-k)G\b] \b)
        \B)^\rho  \\
        & = & \exp_2(-d\rho R + (d-k)s \rho G) \B(\sum_{\wt{u}_1^d} \b
        (\frac{Q(\wt{u}_1^k|v_1^k)}{Q(u_1^k|v_1^k)} \b )^s Q(\wt{u}_{k+1}^d|v_{k+1}^d)^s \B
        )^\rho \\
        & \stackrel{(b)}{=} &  \exp_2(-d\rho R + (d-k)s \rho G) \B(\sum_{\wt{u}_1^k} \b
        (\frac{Q(\wt{u}_1^k|v_1^k)}{Q(u_1^k|v_1^k)} \b )^s \B )^\rho \B(\sum_{\wt{u}_{k+1}^d}
        Q(\wt{u}_{k+1}^d|v_{k+1}^d)^s \B )^\rho
    \end{eqnarray}

Relation (b) follows from the standard algebra of interchanging sums and products. Now, we
substitute the last line into (\ref{eqn:scsiFd}).
    \begin{eqnarray}
        P(F_d) & \leq &  \sum_{k=1}^d 2^{-d\rho R + (d-k)s \rho G } \sum_{v_1^d} Q(v_1^d) \sum_{u_1^k} \sum_{u_{k+1}^d}Q(u_1^k|v_1^k) Q(u_{k+1}^d|v_{k+1}^d) \cdot
        \nonumber \\
        & & \hspace{15mm}  \b(\sum_{\wt{u}_1^k} \b (\frac{Q(\wt{u}_1^k|v_1^k)}{Q(u_1^k|v_1^k)} \b )^s
        \b)^\rho \b (
        \sum_{\wt{u}_{k+1}^d} Q(\wt{u}_{k+1}^d|v_{k+1}^d)^s \b)^\rho\\
        & = &  \sum_{k=1}^d 2^{-d\rho R + (d-k)s \rho G} \sum_{v_1^k} Q(v_1^k) \sum_{v_{k+1}^d} Q(v_{k+1}^d) \sum_{u_1^k} Q(u_1^k|v_1^k)^{1-s\rho} \nonumber \\
         & & \hspace{15mm}  \b(\sum_{\wt{u}_1^k} Q(\wt{u}_1^k|v_1^k)^s
        \b)^\rho \sum_{u_{k+1}^d} Q(u_{k+1}^d|v_{k+1}^d) \b (\sum_{\wt{u}_{k+1}^d} Q(\wt{u}_{k+1}^d|v_{k+1}^d)^s \b)^\rho \\
         & = &  \sum_{k=1}^d 2^{-d\rho R + (d-k)s \rho G}  \sum_{v_1^k} Q(v_1^k) \sum_{v_{k+1}^d} Q(v_{k+1}^d) \b(\sum_{u_1^k} Q(u_1^k|v_1^k)^{1-s\rho}\b) \cdot \nonumber \\
         & & \hspace{1.5cm}\b(\sum_{\wt{u}_1^k} Q(\wt{u}_1^k|v_1^k)^s
        \b)^\rho \b (\sum_{\wt{u}_{k+1}^d} Q(\wt{u}_{k+1}^d|v_{k+1}^d)^s \b)^\rho \\
        & \stackrel{(c)}{=} & \sum_{k=1}^d 2^{-d\rho R + (d-k)\frac{\rho}{1+\rho} G} \sum_{v_1^k}
        Q(v_1^k) \b(\sum_{u_1^k} Q(u_1^k|v_1^k)^{\frac{1}{1+\rho}}\b)^{1+\rho} \cdot \nonumber \\
        & &\hspace{1.5cm} \sum_{v_{k+1}^d} Q(v_{k+1}^d) \b (\sum_{u_{k+1}^d}
        Q(u_{k+1}^d|v_{k+1}^d)^\frac{1}{1+\rho}\b)^\rho
    \end{eqnarray}

We get (c) by noting that the $\wt{u}$'s are just dummy variables and we are free to replace
them with $u$'s and then setting $s = \frac{1}{1+\rho}$. Next, we use the IID property of the
source along with some algebra to get to an exponential form. For example, we have
    \begin{eqnarray}
        \sum_{v_1^k} Q(v_1^k)\b( \sum_{u_1^k} Q(u_1^k|v_1^k)^{\frac{1}{1+\rho}}\b)^{1+\rho} & = & \sum_{v_1} \cdots \sum_{v_k} \prod_{l=1}^k Q(v_l)\b(\sum_{u_1} \sum_{u_2} \cdots \sum_{u_k}
         \prod_{l=1}^k Q(u_l|v_l)^{\frac{1}{1+\rho}} \b)^{1+\rho}\\
        & = & \prod_{l=1}^k \sum_{v_l} Q(v_l)\b(\sum_{u_l} Q(u_l|v_l)^{\frac{1}{1+\rho}}\b)^{1+\rho} \\
        & = & \B[\sum_{v\in \mV} \b( \sum_{u\in \mU} Q(u,v)^{\frac{1}{1+\rho}}
        \b)^{1+\rho}\B]^k
    \end{eqnarray}

Similarly,
    \begin{eqnarray}
        \sum_{v_1^k} Q(v_1^k) \b( \sum_{u_1^k} Q(u_1^k|v_1^k)^\frac{1}{1+\rho} \b)^\rho & = & \sum_{v_1} \cdots \sum_{v_k} \prod_{l=1}^k Q(v_l) \b( \sum_{u_1} \sum_{u_2} \cdots \sum_{u_k}
        \prod_{l=1}^k
        Q(u_l|v_l)^\frac{1}{1+\rho} \b)^\rho \\
        & = &  \prod_{l=1}^k  \sum_{v_l} Q(v_l) \b( \sum_{u_l} Q(u_l|v_l)^\frac{1}{1+\rho} \b)^\rho \\
        & = & \B[ \sum_{v\in \mV} Q(v) \b( \sum_{u\in \mU} Q(u|v)^\frac{1}{1+\rho} \b)^{\rho}\B]^k
    \end{eqnarray}

Using the definitions of $E_{si}(\rho)$ and $F_{si}(\rho)$, we can rewrite the bound as:
    \begin{eqnarray}
        P(F_d)& \leq & \sum_{k=1}^d \exp\b( - d\rho R + (d-k)\frac{\rho}{1+\rho} G + k E_{si}(\rho) + (d-k) F_{si}(\rho) \b)  \\
        & = & \exp\b(d\big ( \frac{\rho}{1+\rho} G + F_{si}(\rho) - \rho R \big) \b) \sum_{k=1}^d \exp
        \b( k \big( E_{si}(\rho) - F_{si}(\rho) - \frac{\rho}{1+\rho} G \big)
    \end{eqnarray}

By the assumption of the theorem, (\ref{eqn:biascondsi}), the following condition holds
    \begin{equation} E_{si}(\rho) - F_{si}(\rho) - \frac{\rho}{1+\rho} G \geq  0
        \label{eqn:biascondition}\end{equation}

Then, we can simplify the bound to
    \begin{eqnarray}
        P(F_d) & \leq & \exp\b(d\big ( \frac{\rho}{1+\rho}G + F_{si}(\rho) - \rho R \big) \b) \sum_{k=1}^d \exp
        \b( k \big( E_{si}(\rho) - F_{si}(\rho) - \frac{\rho}{1+\rho} G \big) \b) \\ & \stackrel{(d)}{\leq} &
        \exp\b(d\big ( \frac{\rho}{1+\rho} G + F_{si}(\rho) - \rho R \big) \b)\cdot d \exp \b( d \big(
        E_{si}(\rho) - F_{si}(\rho) - \frac{\rho}{1+\rho} G \big) \b)
        \\ & = & d \exp \b( -d(\rho R - E_{si}(\rho) )\b)
    \end{eqnarray}

We get (d) from noting that the sum of the geometric series can be upper bounded by $d$ times
the largest term. Now this holds for all $\rho \in [0,1]$, so
    \begin{eqnarray}
        P(F_d) & \leq & K_\eps \exp \b( -d(\rho R - E_{si}(\rho) - \epsilon )
        \b) \\
        \wt{K}_\eps & \triangleq & \max \b \brl d : \frac{\log d}{d} \geq \eps \b \brr < \infty
    \end{eqnarray}

Note that $\wt{K}_\eps < \infty$ and is independent of $d$ because $\ln (d) / d$ goes to $0$.
We note that $E_{si}(\rho)$ is a differentiable function for all $\rho \geq 0$, with
$E_{si}'(0) = H(U|V)$ (see \cite{GallagerSideInformation}); that is, the slope at $0$ is the
conditional entropy of the source $U$ given the side information $V$. $E_{si}(\rho)$ is the
source coding  with side information coding analog to Gallager's function $E_0(\rho)$. While
Gallager's function may be non-differentiable at points because it is the maximization of a
function over probability distributions, $E_{si}(\rho)$ doesn't suffer from this problem.

Now, assuming the bias satisfies the required condition, we have
    \begin{eqnarray}
        P_e(d) & \leq & \sum_{k=0}^{\infty} P(F_{d+k}) \\
        & \leq & \sum_{k=0}^\infty \wt{K}_\eps 2^{-(d+k)(\rho R - E_{si}(\rho) - \eps)} \\
        & = & \exp\b(-d(\rho R - E_{si}(\rho) - \eps)\b) \sum_{k = 0}^\infty \wt{K}_\eps 2^{-k(\rho R - E_{si}(\rho) - \eps)}
    \end{eqnarray}

Since we can choose $\eps$ arbitrarily small, the geometric series converges and we have
    \begin{eqnarray}
        P_e(d) & \leq & \frac{\wt{K}_\eps}{1 - 2^{-(\rho R - E_{si}(\rho) - \eps)}} 2^{-d(\rho R - E_{si}(\rho) - \eps)}\\
        & = & \tilde{K}_\eps \exp_2\b(-d(\rho R - E_{si}(\rho) - \eps)\b)
    \end{eqnarray}

This is true for all $\rho \in [0,1]$, so $E(R) \geq E_{r,si}(R) = \sup_{\rho \in [0,1]} \rho R
- E_{si}(\rho)$.
\end{proof}

\subsection{Probability of error - joint source channel coding with side information}
\label{sec:proberrorjsc}

\begin{theorem}[Restatement of Theorem \ref{thm:proberrorsc}] Suppose there is a channel $W$ between the encoder and the decoder and
side information is available to the decoder. Fix any $\eps > 0$ and let $\rho \in [0,1]$. For
the encoder/decoder of Sections \ref{sec:binningscheme} and \ref{sec:jointscwithsisetup}, if
the bias $G$ satisfies
\begin{equation} G \leq \frac{1+\rho}{\rho}\B[ E_{si}(\rho) - F_{si}(\rho) - \lambda E_0(\rho) + \lambda F(\rho) \B]
\end{equation}
then, there is a constant $K_\eps < \infty$ so that
\begin{equation} P_e(d) < K_\eps \exp_2 \B(-d\b(\lambda E_0(\rho) - E_{si}(\rho) - \eps \b)\B)
\end{equation}
Hence, with suitable choice of bias, the error exponent with delay can be
\begin{equation} E(\lambda) = E_{r,jscsi}(\lambda) \triangleq \sup_{\rho \in [0,1]} \lambda E_0(\rho) - E_{si}(\rho)
\end{equation} \end{theorem}\vspace{.2in}

\begin{proof} We will prove this for $\lambda = 1$ and then show how the proof changes for other $\lambda$. As in the previous proof, $P_e(d)$ can be bounded by $\sum_{l=0}^\infty P(F_{d+l})$, and
$P(F_d)$ can be bounded by $\sum_{k=1}^d P(F_{d,k})$. So we start by bounding $P(F_{d,k})$.
First condition on the true source sequence, channel inputs and channel outputs.
    \begin{eqnarray}
        P(F_{d,k}) & \leq & \sum_{u_1^d,v_1^d,x_1^d,y_1^d} Q(u_1^d,v_1^d)R(x_1^d)W(y_1^d|x_1^d)E\b[ \b(
            \sum_{\wt{u}_1^d \in \mU^d, \wt{u}_1 \neq u_1} 1(\Gamma(\wt{u}_1^d) \geq \Gamma(u_1^k))
            \b)^\rho\b| x_1^d, y_1^d, u_1^d, v_1^d \b] \label{eqn:Alkjsc} \\
        & \leq & \sum_{u_1^d,v_1^d,x_1^d,y_1^d} Q(u_1^d,v_1^d)R(x_1^d)W(y_1^d|x_1^d)
            \B( E \b[  \sum_{\wt{u}_1^d \in \mU^d, \wt{u}_1 \neq u_1} 1(\Gamma(\wt{u}_1^d) \geq
            \Gamma(u_1^k)) \b| x_1^d, y_1^d, u_1^d, v_1^d \b] \B)^\rho
    \end{eqnarray}

The last step is true by Jensen's inequality. Now the only thing that is random in the
expectation is the channel symbols used on the false $\wt{u}$ paths.
    \begin{eqnarray}
        P(F_{d,k}) & \leq & \sum_{u_1^d,v_1^d,x_1^d,~y_1^d} Q(u_1^d,v_1^d)R(x_1^d)W(y_1^d|x_1^d) \cdot
        \\ & & \hspace{2cm} \cdot \B( \sum_{\wt{u}_1^d \in \mU^d,~ \wt{u}_1 \neq u_1} \sum_{\wt{x}_1^d} R(\wt{x}_1^d) E \b [
        1(\Gamma(\wt{u}_1^d) \geq \Gamma(u_1^k)) \b| x_1^d, y_1^d, u_1^d, v_1^d, \wt{x}_1^d \b]
        \B)^\rho
    \end{eqnarray}

Now, we also have for all $s \geq 0$, $ 1(\Gamma(\wt{u}_1^d) \geq \Gamma(u_1^k)) \leq
\exp_2(s(\Gamma(\wt{u}_1^d)-\Gamma(u_1^k)))$. So,
    \begin{eqnarray}
        \Gamma(\wt{u}_1^d) - \Gamma(u_1^k) & = &
        \log_2\frac{Q(\wt{u}_1^d|v_1^d)W(y_1^d|\wt{x}_1^d)}{P(y_1^d)} -
        \log_2\frac{Q(u_1^k|v_1^k)W(y_1^k|x_1^k)}{P(y_1^k)} + (d-k)G \\
        1(\Gamma(\wt{u}_1^d) \geq \Gamma(u_1^k)) & \leq & \h(
        \frac{Q(\wt{u}_1^k|v_1^k)W(y_1^k|\wt{x}_1^k)W(y_{k+1}^d|\wt{x}_{k+1}^d)Q(\wt{u}_{k+1}^d|v_{k+1}^d)}{Q(u_1^k|v_1^k)W(y_1^k|x_1^k)P(y_{k+1}^d)}
        \h)^s 2^{s(d-k)G}
    \end{eqnarray}

We can substitute this expression into the inequality for $P(F_{d,k})$.
    \begin{eqnarray}
        P(F_{d,k}) & \leq & \sum_{u_1^d,v_1^d,x_1^d,y_1^d} Q(v_1^d)
        Q(u_1^k|v_1^k)^{1-s\rho}Q(u_{k+1}^d|v_{k+1}^d)
        R(x_1^k)W(y_1^k|x_1^k)^{1-s\rho}R(x_{k+1}^d)W(y_{k+1}^d|x_{k+1}^d) \cdot
        \\& & \hspace{1cm} \cdot \B( \sum_{\wt{u}_1^d \in \mU^d,~ \wt{u}_1 \neq u_1} \sum_{\wt{x}_1^d}
        R(\wt{x}_1^d)Q(\wt{u}_1^k|v_1^k)^s
        Q(\wt{u}_{k+1}^d|v_{k+1}^d)^sW(y_1^k|\wt{x}_1^k)^s\frac{W(y_{k+1}^d|\wt{x}_{k+1}^d)^s}{P(y_{k+1}^d)^s}
        \B)^\rho 2^{s\rho (d-k)G} \nonumber \\
        & \leq & \sum_{u_1^d,v_1^d,x_1^d,y_1^d} Q(v_1^d)
        Q(u_1^k|v_1^k)^{1-s\rho}Q(u_{k+1}^d|v_{k+1}^d)
        R(x_1^k)W(y_1^k|x_1^k)^{1-s\rho}R(x_{k+1}^d)W(y_{k+1}^d|x_{k+1}^d) \cdot \nonumber
        \\ & & \hspace{1cm} \cdot \B( \sum_{\wt{x}_1^d} R(\wt{x}_1^d) W(y_1^k|\wt{x}_1^k)^s\frac{W(y_{k+1}^d|\wt{x}_{k+1}^d)^s}{P(y_{k+1}^d)^s} \sum_{\wt{u}_1^d}
        Q(\wt{u}_1^k|v_1^k)^s Q(\wt{u}_{k+1}^d|v_{k+1}^d)^s \B)^\rho 2^{s\rho (d-k)G}
    \end{eqnarray}

Now set $s = 1/(1+\rho)$.
    \begin{eqnarray}
        P(F_{d,k}) & \leq & \sum_{u_1^d,v_1^d,x_1^d,y_1^d} \B(Q(v_1^d)
        Q(u_1^k|v_1^k)^\frac{1}{1+\rho} Q(u_{k+1}^d|v_{k+1}^d)\B) \cdot \nonumber
        \\ & & \hspace{1cm} \cdot \B( R(x_1^k)W(y_1^k|x_1^k)^\frac{1}{1+\rho} R(x_{k+1}^d)W(y_{k+1}^d|x_{k+1}^d)\B) \B( \sum_{\wt{x}_1^d} R(\wt{x}_1^d) W(y_1^k|\wt{x}_1^k)^\frac{1}{1+\rho} \frac{W(y_{k+1}^d|\wt{x}_{k+1}^d)^\frac{1}{1+\rho} }{P(y_{k+1}^d)^\frac{1}{1+\rho} } \B)^\rho \nonumber \\
        & & \hspace{2cm} \B(\sum_{\wt{u}_1^d} Q(\wt{u}_1^k|v_1^k)^\frac{1}{1+\rho}  Q(\wt{u}_{k+1}^d|v_{k+1}^d)^\frac{1}{1+\rho} \B)^\rho 2^{\frac{\rho}{1+\rho} (d-k)G}\\
         & \leq & \B(\sum_{u_1^d,v_1^d}  \B(Q(v_1^d) Q(u_1^k|v_1^k)^\frac{1}{1+\rho} Q(u_{k+1}^d|v_{k+1}^d)\B) \B(\sum_{\wt{u}_1^d}
        Q(\wt{u}_1^k|v_1^k)^\frac{1}{1+\rho} Q(\wt{u}_{k+1}^d|v_{k+1}^d)^\frac{1}{1+\rho} \B)^\rho 2^{\frac{\rho}{1+\rho}(d-k)G} \B)
        \cdot \nonumber \\ & &  \B(\sum_{x_1^d,y_1^d} R(x_1^k)W(y_1^k|x_1^k)^\frac{1}{1+\rho}R(x_{k+1}^d)W(y_{k+1}^d|x_{k+1}^d)
        \B( \sum_{\wt{x}_1^d} R(\wt{x}_1^d)
        W(y_1^k|\wt{x}_1^k)^\frac{1}{1+\rho}\frac{W(y_{k+1}^d|\wt{x}_{k+1}^d)^\frac{1}{1+\rho}}{P(y_{k+1}^d)^\frac{1}{1+\rho}} \B)^\rho
    \end{eqnarray}

To further reduce this expression, notice $P(F_{d,k}) \leq A \cdot B\cdot
2^{\frac{\rho}{1+\rho} (d-k)G}$, where
    \begin{eqnarray}
        A & \triangleq & \sum_{u_1^d,v_1^d}  \B(Q(v_1^d) Q(u_1^k|v_1^k)^\frac{1}{1+\rho} Q(u_{k+1}^d|v_{k+1}^d)\B)
        \B(\sum_{\wt{u}_1^d} Q(\wt{u}_1^k|v_1^k)^\frac{1}{1+\rho} Q(\wt{u}_{k+1}^d|v_{k+1}^d)^\frac{1}{1+\rho} \B)^\rho \\
        B &\triangleq & \sum_{x_1^d,y_1^d} R(x_1^k)W(y_1^k|x_1^k)^\frac{1}{1+\rho} R(x_{k+1}^d)W(y_{k+1}^d|x_{k+1}^d)
        \B( \sum_{\wt{x}_1^d} R(\wt{x}_1^d)
        W(y_1^k|\wt{x}_1^k)^\frac{1}{1+\rho}\frac{W(y_{k+1}^d|\wt{x}_{k+1}^d)^\frac{1}{1+\rho}}{P(y_{k+1}^d)^\frac{1}{1+\rho}} \B)^\rho
    \end{eqnarray}

Now, we work on each term individually. $A$ can be written in two parts, $A = A_1 \cdot A_2$,
where $A_1$ is the term corresponding to the letters from time $1$ to $k$ and $A_2$ is the term
corresponding to letters from time $k+1$ to $d$. Explanations for steps are given after the
equations.
    \begin{eqnarray}
        A_1 & \triangleq & \sum_{v_1^k} Q(v_1^k)\sum_{u_1^k}  Q(u_1^k|v_1^k)^\frac{1}{1+\rho} \B ( \sum_{\wt{u}_1^k}
        Q(\wt{u}_1^k|v_1^k)^\frac{1}{1+\rho} \B)^\rho \\
        & \sr{(a)}{=} & \sum_{v_1} \cdots \sum_{v_k}Q(v_1^k) \sum_{u_1} \cdots \sum_{u_k} \prod_{l=1}^k
        Q(u_l|v_l)^\frac{1}{1+\rho}
        \B( \sum_{\wt{u}_1} \cdots \sum_{\wt{u}_k} \prod_{j=1}^k Q(\wt{u}_j|v_j)^\frac{1}{1+\rho} \B)^\rho \\
        & \sr{(b)}{=}& \sum_{v_1} \cdots \sum_{v_k}Q(v_1^k) \sum_{u_1} \cdots \sum_{u_k} \prod_{l=1}^k
        Q(u_l|v_l)^\frac{1}{1+\rho}
        \B( \prod_{j=1}^k \sum_{\wt{u}_j} Q(\wt{u}_j|v_j)^\frac{1}{1+\rho} \B)^\rho \\
        & \sr{(c)}{=} & \sum_{v_1}\cdots \sum_{v_k}Q(v_1^k) \B( \prod_{l=1}^k \sum_{u_l}  Q(u_l|v_l)^\frac{1}{1+\rho} \B) \B(
        \prod_{j=1}^k \b(\sum_{\wt{u}_j} Q(\wt{u}_j|v_j)^\frac{1}{1+\rho} \b)^\rho \B) \\
        & \sr{(d)}{=} & \sum_{v_1}\cdots \sum_{v_k}Q(v_1^k) \B( \prod_{l=1}^k \sum_{u_l}  Q(u_l|v_l)^\frac{1}{1+\rho} \B) \B(
        \prod_{j=1}^k \b(\sum_{u_j} Q(u_j|v_j)^\frac{1}{1+\rho} \b)^\rho \B) \\
        & \sr{(e)}{=} & \sum_{v_1} \cdots \sum_{v_k} Q(v_1^k) \prod_{l=1}^k \B( \sum_{u_l}
        Q(u_l|v_l)^{\frac{1}{1+\rho}} \B)^{1+\rho} \\
        & \sr{(f)}{=} & \sum_{v_1}\cdots \sum_{v_k} \prod_{l=1}^k Q(v_l) \B( \sum_{u_l}
        Q(u_l|v_l)^{\frac{1}{1+\rho}} \B)^{1+\rho} \\
        & \sr{(g)}{=} &  \prod_{l=1}^k \h( \sum_{v_l} Q(v_l) \B( \sum_{u_l} Q(u_l|v_l)^{\frac{1}{1+\rho}}
        \B)^{1+\rho} \h)\\
        & \sr{(h)}{=} & \B( \sum_{v \in \mV} Q(v) \b( \sum_{u \in \mU} Q(u|v)^\frac{1}{1+\rho} \b)^{1+\rho} \B)^k
        \\
        & = & \exp_2\B( k \log_2 \B[\sum_{v\in \mV} Q(v) \b(\sum_{u\in \mU} Q(u|v)^\frac{1}{1+\rho}
        \b)^{1+\rho}\B]\B)
    \end{eqnarray}
    \begin{enumerate}
        \item[a)] Memorylessness of source.
        \item[b)] Sums and products commute.
        \item[c)] Same as last step.
        \item[d)] Replace dummy variables.
        \item[e)] Combine common terms.
        \item[f)] Memorylessness of source.
        \item[g)] Commuting sum and product.
        \item[h)] Dummy variable replacement, each of the $k$ terms is the same.
    \end{enumerate}

Similarly, we work out $A_2$ below.
    \begin{eqnarray}
         A_2 & \triangleq & \sum_{v_{k+1}^d} Q(v_{k+1}^d) \sum_{u_{k+1}^d} Q(u_{k+1}^d|v_{k+1}^d) \b(
         \sum_{\wt{u}_{k+1}^d} Q(\wt{u}_{k+1}^d|v_{k+1}^d)^\frac{1}{1+\rho} \b)^\rho \label{eqn:jscsiC}\\
         & = & \sum_{v_{k+1}^d} Q(v_{k+1}^d) \b(\sum_{u_{k+1}^d} Q(u_{k+1}^d|v_{k+1}^d)\b) \b(
         \sum_{\wt{u}_{k+1}^d} Q(\wt{u}_{k+1}^d|v_{k+1}^d)^\frac{1}{1+\rho} \b)^\rho \\
         & \sr{(a)}{=} & \sum_{v_{k+1}^d} Q(v_{k+1}^d) \cdot 1 \cdot \b(\sum_{\wt{u}_{k+1}^d}
         Q(\wt{u}_{k+1}^d|v_{k+1}^d)^\frac{1}{1+\rho} \b)^\rho \\
         & \sr{(b)}{=} &  \sum_{v_{k+1}^d} Q(v_{k+1}^d) \b(\sum_{u_{k+1}^d}
         Q(u_{k+1}^d|v_{k+1}^d)^\frac{1}{1+\rho} \b)^\rho \\
         & \sr{(c)}{=} & \prod_{l=k+1}^d \sum_{v_l} Q(v_l) \b( \sum_{u_l} Q(u_l|v_l)^\frac{1}{1+\rho} \b)^\rho \\
         & \sr{(d)}{=} & \B( \sum_{v} Q(v) \b( \sum_{u} Q(u|v)^\frac{1}{1+\rho} \b)^\rho\B)^{d-k} \\
         & = & \exp_2 \B( (d-k) \log_2 \B[\sum_{v\in \mV} Q(v) \b( \sum_{u\in \mU} Q(u|v)^\frac{1}{1+\rho}
         \b)^\rho\B ] \B)
    \end{eqnarray}
    \begin{itemize}
        \item[a)] The sum of the probabilities in a conditional distribution is $1$.
        \item[b)] Replace dummy variable.
        \item[c)] Memorylessness of source.
        \item[d)] All $d-k$ terms in the product are the same.
    \end{itemize}

Now use the definitions of $E_{si}$ and $F_{si}$ to write $A$ as:
    \begin{eqnarray}
        A & = & A_1 \cdot A_2 \\
        & = & \exp_2 \b( k E_{si}(\rho) + (d-k) F_{si}(\rho) \b)
    \end{eqnarray}

Analogously, we will write $B = B_1\cdot B_2$ where $B_1$ is the product of terms concerning
time $1$ to $k$ and $B_2$ is the product of terms concerning time $k+1$ to $d$.
    \begin{eqnarray}
        B_1 & \triangleq & \sum_{y_1^k} \b(\sum_{x_1^k} R(x_1^k) W(y_1^k|x_1^k)^{\frac{1}{1+\rho}} \b)\b(
        \sum_{\wt{x}_1^k} R(\wt{x}_1^k) W(y_1^k|\wt{x}_1^k)^\frac{1}{1+\rho} \b)^\rho \\
         & \sr{(a)}{=} & \sum_{y_1^k} \b( \sum_{x_1^k} R(x_1^k) W(y_1^k|x_1^k)^{\frac{1}{1+\rho}} \b)^{1+\rho} \\
         & \sr{(b)}{=} & \prod_{l=1}^k \sum_{y_l} \b( \sum_{x_l} R(x_l)W(y_l|x_l)^\frac{1}{1+\rho} \b)^{1+\rho}\\
         & \sr{(c)}{=} & \B(\sum_{y\in \mY} \b( \sum_{x\in \mX} \Beta(x)W(y|x)^{\frac{1}{1+\rho}} \b)^{1+\rho}\B)^k
         \\
         & = & \exp_2 \B( k \log_2 \b[ \sum_{y\in\mY} \b( \sum_{x\in \mX}
         \Beta(x)W(y|x)^\frac{1}{1+\rho}\b)^{1+\rho} \b] \B)
    \end{eqnarray}
    \begin{itemize}
        \item[a)] Replace dummy variables and combine common terms.
        \item[b)] Use source memorylessness and commute products with sums.
        \item[c)] All $k$ terms in the product are the same, replace the dummy variables.
    \end{itemize}

Similarly for $B_2$,
    \begin{eqnarray}
        B_2 & \triangleq & \sum_{y_{k+1}^d} \B(\sum_{x_{k+1}^d} R(x_{k+1}^d) W(y_{k+1}^d|x_{k+1}^d)\B)
            \B( \sum_{\wt{x}_{k+1}^d} R(\wt{x}_{k+1}^d) \B[
            \frac{W(y_{k+1}^d|\wt{x}_{k+1}^d)}{P(y_{k+1}^d)}\B]^\frac{1}{1+\rho}\B)^\rho \\
        & \sr{(a)}{=} & \sum_{y_{k+1}^d} P(y_{k+1}^d) \B( \sum_{\wt{x}_{k+1}^d} R(\wt{x}_{k+1}^d) \B[
            \frac{W(y_{k+1}^d|\wt{x}_{k+1}^d)}{P(y_{k+1}^d)}\B]^\frac{1}{1+\rho}\B)^\rho \\
        & \sr{(b)}{=} & \sum_{y_{k+1}^d} P(y_{k+1}^d) \B( \sum_{x_{k+1}^d} R(x_{k+1}^d) \B[
            \frac{W(y_{k+1}^d|x_{k+1}^d)}{P(y_{k+1}^d)}\B]^\frac{1}{1+\rho}\B)^\rho \\
        & \sr{(c)}{=} & \sum_{y_{k+1}^d} P(y_{k+1}^d)^\frac{1}{1+\rho}\B( \sum_{x_{k+1}^d} R(x_{k+1}^d)
            W(y_{k+1}^d|x_{k+1}^d)^\frac{1}{1+\rho} \B)^\rho \\
        & \sr{(d)}{=} & \prod_{l=k+1}^d \sum_{y_l}P(y_l)^\frac{1}{1+\rho} \B(\sum_{x_l} R(x_l)
            W(y_l|x_l)^\frac{1}{1+\rho}\B)^\rho \\
        & \sr{(e)}{=} & \B( \sum_{y}P(y)^\frac{1}{1+\rho} \B(\sum_{x} \Beta(x)
            W(y|x)^\frac{1}{1+\rho}\B)^\rho\B)^{d-k} \\
        & = & \exp_2 \h( (d-k) \log_2 \h[ \sum_{y\in \mY}P(y) \B( \sum_{x\in \mX} \Beta(x) \B[
            \frac{W(y|x)}{P(y)} \B]^\frac{1}{1+\rho} \B)^\rho\h] \h)
    \end{eqnarray}
    \begin{itemize}
        \item[a)] Total probability: the sum in the first parentheses equals $P(y_{k+1}^d)$.
        \item[b)] Replace dummy variables.
        \item[c)] Move $P(y_{k+1}^d)$ out of second sum.
        \item[d)] Memorylessness of channel, IID channel input generation and commute product
        with sums.
        \item[e)] All $d-k$ terms are the same; replace dummy variables.
    \end{itemize}

Use the definitions of $E_0$ and $F$ and substitute for $B_1$ and $B_2$ to get:
    \begin{eqnarray}
        B & = & B_1\cdot B_2 \\
         & = & \exp_2\b(-kE_0(\rho) - (d-k)F(\rho)\b)
    \end{eqnarray}

Finally, we can put everything together:
    \begin{eqnarray}
        P(F_{d,k}) & \leq & A\cdot B \\
        & = & \exp_2 \h( (d-k) \frac{\rho}{1+\rho} G + k E_{si}(\rho) + (d-k) F_{si}(\rho) - k
        E_0(\rho) - (d-k) F(\rho) \h) \\
        P(F_d) & \leq & \sum_{k=1}^d P(F_{d,k}) \\
        & \leq & \sum_{k=1}^d \exp_2 \h( (d-k) \frac{\rho}{1+\rho} G + k E_{si}(\rho) + (d-k) F_{si}(\rho) - k
        E_0(\rho) - (d-k) F(\rho) \h)\\
        & = & \exp_2\h(d\B[ \frac{\rho}{1+\rho}G + F_{si}(\rho) - F(\rho)\B]\h) \cdot\nonumber \\
        & & \hspace{2cm} \cdot \sum_{k=1}^d \exp_2\h( k \B[ E_{si}(\rho)-F_{si}(\rho) - E_0(\rho) +
        F(\rho) - \frac{\rho}{1+\rho}G \B] \h)
    \end{eqnarray}

{\em Now suppose that $\lambda \neq 1$}. The only thing that would change would be that instead
of $d$ channel inputs and outputs, there would be $\lambda d$ channel inputs and outputs. The
independence of the channel and source straightforwardly gives:
    \begin{eqnarray}
        P(F_{d,k}) & \leq & \exp_2 \h( (d-k) \frac{\rho}{1+\rho} G + k E_{si}(\rho) + (d-k) F_{si}(\rho) -
        k\lambda E_0(\rho) - (d-k)\lambda F(\rho) \h) \\
        P(F_d) & \leq & \exp_2\h(d\B[ \frac{\rho}{1+\rho}G + F_{si}(\rho) - \lambda F(\rho)\B]\h) \cdot \nonumber \\
        & & \hspace{2cm} \cdot \sum_{k=1}^d \exp_2\h( k \B[ E_{si}(\rho)-F_{si}(\rho) - \lambda E_0(\rho) +
        \lambda F(\rho) - \frac{\rho}{1+\rho}G \B] \h)
    \end{eqnarray}

Now, we assume that $ G \leq \frac{1+\rho}{\rho} [ E_{si}(\rho) - F_{si}(\rho) - \lambda
E_0(\rho) + \lambda F(\rho) ]$, so that the term in the exponential in the sum is positive.
Then the total sum can be bounded by $d$ times the $d^{th}$ term in the sum.
    \begin{eqnarray}
        P(F_d) & \leq & d \exp\h(-d \B( \lambda E_0(\rho) - E_{si}(\rho) \B) \h)
    \end{eqnarray}

The derivative at zero of $E_0$ is $I(R,W)$ where
    \begin{equation} I(R,W) = \sum_{x\in \mX, y \in \mY} \Beta(x)W(y|x) \log_2 \frac{\Beta(x)W(y|x)}{\Beta(x)P(y)} \end{equation}
and the derivative of $E_{si}$ at zero is $H(U|V)$, so if $H(U|V) < \lambda I(R,W)$, there is
some $\rho \in (0,1]$ so that the difference $\lambda E_0(\rho) - E_{si}(\rho)$ is strictly
positive. The $\rho$ can be optimized to give the source-channel random coding with side
information exponent $E_{r,sc}(\lambda) = \max_{\rho \in [0,1]} \lambda E_0(\rho) -
E_{si}(\rho)$.
\end{proof}

\subsection{Random variable of computation - source coding with side information} \label{sec:compsi}

\begin{theorem}[Restatement of Theorem \ref{thm:compsi}] Suppose that the decoder has access to the side
information and there is a rate $R$ noiseless, binary channel between the encoder and decoder.
Fix any $\gamma \in [0,1]$. For the encoder/decoder of Section \ref{sec:binningscheme}, if the
bias $G$ satisfies
\begin{equation} \frac{1+\gamma}{\gamma}G_{si}(\gamma) < G <
\frac{1+\gamma}{\gamma} \B[ \gamma R - F_{si}(\gamma)\B] \end{equation} then the $\gamma^{th}$
moment of computation is uniformly finite all for $i$, i.e. $\exists ~K < \infty$ such that
$\forall~ i,~ E[N_i^\gamma] < K$, if
\begin{equation}
R > \frac{E_{si}(\gamma)}{\gamma} \end{equation} \end{theorem} \vspace{.2in}

\begin{proof} Recall that
    \begin{eqnarray}
        E[N^\gamma] & \leq & \sum_{l=1}^\infty \sum_{k=1}^\infty
        A_{l,k} \\
        A_{l,k} & \triangleq & E \h[ \B( \sum_{\wt{u}_1^l:~ \wt{u}_1 \neq u_1} 1\b(\Gamma(\wt{u}_1^l)
        \geq \Gamma(u_1^k)\b) \B)^\gamma\h]
    \end{eqnarray}

From section \ref{sec:proberrorsi},(\ref{eqn:Alkscsi}), we already know that if $l \geq k$,
    \begin{equation}
        A_{l,k} \leq \exp_2 \B( (l-k)\frac{\gamma}{1+\gamma}G + k E_{si}(\gamma) + (l-k) F_{si}(\gamma) -
        l \gamma R \B)
    \end{equation}

If $l \leq k$, we have
    \begin{eqnarray}
        A_{l,k} & = & \sum_{u_1^k,v_1^k} Q(u_1^k, v_1^k) E \h[ \B( \sum_{\wt{u}_1^l:\wt{u}_1 \neq u_1}
        1(\Gamma(\wt{u}_1^l) \geq \Gamma(u_1^k)) \B)^\gamma \h| u_1^k, v_1^k \h] \\
        & \stackrel{(a)}{\leq} & \sum_{u_1^k,v_1^k} Q(u_1^k, v_1^k) \B( \sum_{\wt{u}_1^l:\wt{u}_1 \neq
        u_1} E \b[ 1(\Gamma(\wt{u}_1^l) \geq \Gamma(u_1^k))  \b| u_1^k, v_1^k \b] \B)^\gamma \\
        & \leq & \sum_{u_1^k,v_1^k} Q(u_1^k, v_1^k) \B( \sum_{\wt{u}_1^l:\wt{u}_1 \neq u_1} E \b[
        1(\textrm{parities of $\wt{u}_1^l$ match}) \exp_2 \b( \frac{\Gamma(\wt{u}_1^l) -
        \Gamma(u_1^k)}{1+\gamma}\b) \b| u_1^k, v_1^k \b] \B)^\gamma
    \end{eqnarray}

(a) uses Jensen's inequality followed by linearity of conditional expectation. The parity
generation process is independent on different branches of the encoding tree, and
    \begin{eqnarray}
        \exp_2(\Gamma(\wt{u}_1^l) - \Gamma(u_1^k) )& = & \B(\frac{Q(\wt{u}_1^l|v_1^l)}{Q(u_1^l|v_1^l)
        Q(u_{l+1}^k|v_{l+1}^k)} \B)2^{(l-k)G}
    \end{eqnarray}

so substituting gives
    \begin{eqnarray}
        A_{l,k} & \leq & \sum_{u_1^k,v_1^k} Q(u_1^k, v_1^k) \B( \sum_{\wt{u}_1^l:\wt{u}_1 \neq u_1}
        \B(\frac{Q(\wt{u}_1^l|v_1^l)}{Q(u_1^l|v_1^l) Q(u_{l+1}^k|v_{l+1}^k)} \B)^\frac{1}{1+\gamma}
        2^{\frac{(l-k)G}{1+ \gamma} - l R} \B)^\gamma \\
        & \leq & \sum_{u_1^k,v_1^k}
        Q(v_1^l)Q(u_1^l|v_1^l)^\frac{1}{1+\gamma}Q(v_{l+1}^k)Q(u_{l+1}^k|v_{l+1}^k)^\frac{1}{1+\gamma}
        \cdot \nonumber \\
        & & \hspace{2cm} \cdot \exp_2 \B ( (l-k)\frac{\gamma}{1+\gamma}G - l\gamma R \B ) \h(
        \sum_{\wt{u}_1^l} Q(\wt{u}_1^l|v_1^l)^\frac{1}{1+\gamma}\h)^\gamma \\
        & = & \exp_2\B((l-k)\frac{\gamma}{1+\gamma}G - l\gamma R \B ) \cdot C \cdot D\\
        C & \triangleq & \sum_{u_1^l, v_1^l} Q(v_1^l)Q(u_1^l|v_1^l)^\frac{1}{1+\gamma} \h(
        \sum_{\wt{u}_1^l} Q(\wt{u}_1^l|v_1^l)^\frac{1}{1+\gamma}\h)^\gamma \\
        D & \triangleq & \sum_{u_{l+1}^k, v_{l+1}^k}
        Q(v_{l+1}^k)Q(u_{l+1}^k|v_{l+1}^k)^\frac{1}{1+\gamma}
    \end{eqnarray}

The terms corresponding to letters from time $1$ to $l$, $C$ are the same as (\ref{eqn:jscsiC})
in section \ref{sec:proberrorsi}, so we have
    \begin{eqnarray}
        A_{l,k} & \leq & \exp\B((l-k)\frac{\gamma}{1+\gamma}G - l\gamma R + l E_{si}(\gamma) \B )
        \cdot D
    \end{eqnarray}

The term $D$ can be simplified into an exponential form using $G_{si}$:
    \begin{eqnarray}
        D & \triangleq &  \sum_{u_{l+1}^k, v_{l+1}^k}
        Q(v_{l+1}^k)Q(u_{l+1}^k|v_{l+1}^k)^\frac{1}{1+\gamma} \\
        & = & \sum_{v_{l+1}^k} Q(v_{l+1}^k) \sum_{u_{l+1}^k} Q(u_{l+1}^k|v_{l+1}^k)^\frac{1}{1+\gamma} \\
        & = & \prod_{m=l+1}^k \sum_{v_m} Q(v_m) \sum_{u_m} Q(u_m|v_m)^\frac{1}{1+\gamma} \\
        & = & \B( \sum_{v\in \mV} Q(v) \sum_{u\in \mV} Q(u|v)^\frac{1}{1+\gamma} \B)^{k-l}\\
        & = & \exp \B( (k-l) G_{si}(\gamma) \B)
    \end{eqnarray}

So if $k \geq l$, we have:
    \begin{eqnarray}
        A_{l,k} & \leq & \exp\B((l-k)\frac{\gamma}{1+\gamma}G - l\gamma R + l E_{si}(\gamma)  + (k-l)
        G_{si}(\gamma)\B )
    \end{eqnarray}

Combining the bounds gives
    \begin{eqnarray}
        E[N^\gamma] & \leq & \sum_{l=1}^\infty \sum_{k=1}^\infty A_{l,k} \\
        & = & \sum_{l=1}^\infty \sum_{k=l}^\infty A_{l,k} + \sum_{k=1}^\infty \sum_{l=k+1}^\infty
            A_{l,k} \\
        & \leq & \sum_{l=1}^\infty \sum_{k=l}^\infty A_{l,k} + \sum_{k=1}^\infty \sum_{l=k}^\infty
            A_{l,k} \\
        & \sr{(a)}{\leq} & \sum_{l=1}^\infty \sum_{k=l}^\infty \exp \B( (l-k) \frac{\gamma}{1+\gamma} G - l\gamma R + l E_{si}(\gamma)  + (k-l)
            G_{si}(\gamma)\B) + \nonumber \\
        & & \hspace{2cm} \sum_{k=1}^\infty \sum_{l=k}^\infty \exp \B( (l-k)\frac{\gamma}{1+\gamma}G +
            k E_{si}(\gamma) + (l-k) F_{si}(\gamma) - l \gamma R \B) \\
        & \sr{(b)}{=} & \sum_{l=1}^\infty \exp\B( -l\b(\gamma R - E_{si}(\gamma) \b)\B)\sum_{k=l}^\infty \exp
            \B( -(k-l)\b( \frac{\gamma}{1+\gamma} G - G_{si}(\gamma)\b)\B) + \nonumber \\
        & & \hspace{1cm} \sum_{k=1}^\infty \exp \B( -k \b( \gamma R - E_{si}(\gamma) \b)
            \B)\sum_{l=k}^\infty \exp \B( -(l-k)\b(-\frac{\gamma}{1+\gamma}G +  F_{si}(\gamma) -
            \gamma R\b)\B)
    \end{eqnarray}
    \begin{itemize}
        \item[a)] Substitute for $A_{l,k}$.
        \item[b)] Add and subtract $(l-k)\gamma R$ in the exponent of the second double sum.
    \end{itemize}

The above sums converge if the following conditions are met:
    \begin{eqnarray}
          \gamma R & > & E_{si}(\gamma) \label{eqn:condone}\\
         \frac{\gamma}{1+\gamma}G & > & G_{si}(\gamma) \label{eqn:condtwo} \\
         \frac{\gamma}{1+\gamma}G & < & \gamma R - F_{si}(\gamma) \label{eqn:condthree}
    \end{eqnarray}

This concludes the proof assuming these conditions hold. To see that (\ref{eqn:condtwo}) and
(\ref{eqn:condthree}) can be satisfied by one choice of bias assuming (\ref{eqn:condone}), see
section \ref{sec:nonemptyrange}.
\end{proof}

\subsection{Random variable of computation - joint source channel coding with side information}
\label{sec:compjsc}

\begin{theorem}[Restatement of Theorem \ref{thm:compsc}] Suppose there is a channel $W$ between the encoder and the decoder and
side information is available to the decoder. Fix any $\gamma \in [0,1]$. For the
encoder/decoder of Sections \ref{sec:binningscheme} and \ref{sec:jointscwithsisetup}, if the
bias $G$ satisfies
\begin{equation} \frac{1+\gamma}{\gamma}\B[ G_{si}(\gamma) - \lambda G(\gamma) \B] < G <
\frac{1+\gamma}{\gamma} \B[\lambda F(\gamma) - F_{si}(\gamma)\B] \end{equation} then the
$\gamma^{th}$ moment of computation is uniformly finite all for $i$, i.e. $\exists ~K < \infty$
such that $\forall~ i,~ E[N_i^\gamma] < K$, if
\begin{equation}
\lambda E_0(\gamma) > E_{si}(\gamma)\end{equation} \end{theorem}\vspace{.2in}

\begin{proof} Again, we will show this for $\lambda = 1$ and at the end see how it changes for $\lambda \neq 1$. Recall that
    \begin{eqnarray}
        E[N^\gamma] & \leq & \sum_{l=1}^\infty \sum_{k=1}^\infty
        A_{l,k} \\
        A_{l,k} & \triangleq & E \h[ \B( \sum_{\wt{u}_1^l:~ \wt{u}_1 \neq u_1} 1\b(\Gamma(\wt{u}_1^l)
        \geq \Gamma(u_1^k)\b) \B)^\gamma \h]
    \end{eqnarray}

From (\ref{eqn:Alkjsc}) in section \ref{sec:proberrorjsc}, we already know that if $l \geq k$,
    \begin{equation}
        A_{l,k} \leq \exp \B( (l-k)\frac{\gamma}{1+\gamma}G + k E_{si}(\gamma) + (l-k) F_{si}(\gamma)
        - k E_0(\gamma) - (l-k) F(\gamma) \B)
    \end{equation}

If $l \leq k$, we have
    \begin{eqnarray}
        A_{l,k} & = & \sum_{u_1^k,v_1^k,x_1^k,y_1^k} Q(u_1^k, v_1^k)R(x_1^k)W(y_1^k|x_1^k) E \h[ \B(
            \sum_{\wt{u}_1^l} 1(\Gamma(\wt{u}_1^l) \geq \Gamma(u_1^k)) \B)^\gamma \h| x_1^k, y_1^k, u_1^k,
            v_1^k \h] \\
        & \sr{(a)}{\leq} & \sum_{u_1^k,v_1^k,x_1^k,y_1^k} Q(u_1^k, v_1^k)R(x_1^k)W(y_1^k|x_1^k) E \h[
            \sum_{\wt{u}_1^l} 1(\Gamma(\wt{u}_1^l) \geq \Gamma(u_1^k))  \h| x_1^k, y_1^k, u_1^k, v_1^k
            \h]^\gamma\\
        & \leq & \sum_{u_1^k,v_1^k,x_1^k,y_1^k} Q(u_1^k, v_1^k)R(x_1^k)W(y_1^k|x_1^k) E \h[
            \sum_{\wt{u}_1^l}\exp\B(\frac{\Gamma(\wt{u}_1^l)-\Gamma(u_1^k)}{1+\gamma}\B)  \h| x_1^k, y_1^k,
            u_1^k, v_1^k \h]^\gamma \\
        & \sr{(b)}{=} & \sum_{u_1^k,v_1^k,x_1^k,y_1^k} Q(u_1^k, v_1^k)R(x_1^k)W(y_1^k|x_1^k) \h(
            \sum_{\wt{u}_1^l} E\B[\exp\B(\frac{\Gamma(\wt{u}_1^l)-\Gamma(u_1^k)}{1+\gamma}\B)  \B| x_1^k,
            y_1^k, u_1^k,v_1^k \B]\h)^\gamma \\
        & \sr{(c)}{=} & \sum_{u_1^k,v_1^k,x_1^k,y_1^k} Q(u_1^k, v_1^k)R(x_1^k)W(y_1^k|x_1^k) \h(
            \sum_{\wt{u}_1^l} \sum_{\wt{x}_1^l} R(\wt{x}_1^l)
            \exp\B(\frac{\Gamma(\wt{u}_1^l)-\Gamma(u_1^k)}{1+\gamma}\B) \h)^\gamma
    \end{eqnarray}
    \begin{itemize}
        \item[a)] Jensen's inequality.
        \item[b)] Linearity of expectation.
        \item[c)] Conditioning on channel inputs along `false' paths.
    \end{itemize}

Now, we write out the term in the exponent to get:
    \begin{eqnarray}
        \exp\B(\frac{\Gamma(\wt{u}_1^l)-\Gamma(u_1^k)}{1+\gamma}\B) & = & \h(
        \frac{Q(\wt{u}_1^l|v_1^l)W(y_1^l|\wt{x}_1^l)P(y_{l+1}^k)}{Q(u_1^l|v_1^l)W(y_1^l|x_1^l)Q(u_{l+1}^k|v_{l+1}^k)W(y_{l+1}^k|x_{l+1}^k)}
        \h)^\frac{1}{1+\gamma} 2^{(l-k) \frac{1}{1+\gamma} G}
    \end{eqnarray}

So, substituting and merging terms gives:
    \begin{eqnarray}
        A_{l,k} & \leq & \sum_{u_1^k,v_1^k,x_1^k,y_1^k}
            Q(v_1^l)Q(u_1^l|v_1^l)^\frac{1}{1+\rho}Q(v_{l+1}^k)Q(u_{l+1}^k|v_{l+1}^k)^\frac{1}{1+\rho}R(x_1^l)W(y_1^l|x_1^l)^{\frac{1}{1+\gamma}}
            \cdot\nonumber  \\
        & & \hspace{2cm} R(x_{l+1}^k)W(y_{l+1}^k|x_{l+1}^k)^\frac{1}{1+\gamma}
            P(y_{l+1}^k)^{\frac{\gamma}{1+\gamma}} 2^{(l-k)\frac{\gamma}{1+\gamma}G} \cdot \nonumber  \\
        & & \hspace{2cm} \h( \sum_{\wt{u}_1^l}  Q(\wt{u}_1^l|v_1^l)^\frac{1}{1+\rho}\h)^\gamma \h( \sum_{\wt{x}_1^l}
            R(\wt{x}_1^l)W(y_1^l|\wt{x}_1^l)^{\frac{1}{1+\gamma}} \h)^\gamma
    \end{eqnarray}

The terms corresponding to letters from time $1$ to $l$ are easily recognized to be the same as
in the last sections, so we can extract them and get
    \begin{eqnarray}
        A_{l,k} & \leq & \exp \h( (l-k) \frac{\gamma}{1+\gamma} G + l E_{si}(\gamma) - l E_0(\gamma)
            \h) \h( \sum_{u_{l+1}^k,v_{l+1}^k} Q(v_{l+1}^k) Q(u_{l+1}^k|v_{l+1}^k)^\frac{1}{1+\gamma} \h) \cdot \nonumber \\
        & & \hspace{3cm} \h( \sum_{x_{l+1}^k,y_{l+1}^k}
            R(x_{l+1}^k)W(y_{l+1}^k|x_{l+1}^k)^\frac{1}{1+\gamma}P(y_{l+1}^k)^\frac{\gamma}{1+\gamma} \h)
            \\
        & = & \exp \B( (l-k) \frac{\gamma}{1+\gamma} G + l E_{si}(\gamma) - l E_0(\gamma) \B) \cdot C
            \cdot D \\
        C & \triangleq &  \sum_{u_{l+1}^k,v_{l+1}^k} Q(v_{l+1}^k)
            Q(u_{l+1}^k|v_{l+1}^k)^\frac{1}{1+\gamma}\\
        D & \triangleq & \sum_{x_{l+1}^k,y_{l+1}^k}
            R(x_{l+1}^k)W(y_{l+1}^k|x_{l+1}^k)^\frac{1}{1+\gamma}P(y_{l+1}^k)^\frac{\gamma}{1+\gamma}
    \end{eqnarray}

Then, we work with $C$ and $D$ individually to get:
    \begin{eqnarray}
        C & = & \sum_{u_{l+1}^k,v_{l+1}^k} Q(v_{l+1}^k) Q(u_{l+1}^k|v_{l+1}^k)^\frac{1}{1+\gamma} \\
        & = & \B( \sum_{v \in \mV} Q(v) \sum_{u \in \mU} Q(u|v)^\frac{1}{1+\gamma}\h)^{k-l} \\
        D & = &  \sum_{x_{l+1}^k,y_{l+1}^k}
            R(x_{l+1}^k)W(y_{l+1}^k|x_{l+1}^k)^\frac{1}{1+\gamma}P(y_{l+1}^k)^\frac{\gamma}{1+\gamma} \\
        & = &  \sum_{x_{l+1}^k,y_{l+1}^k}
            R(x_{l+1}^k)\h(\frac{W(y_{l+1}^k|x_{l+1}^k)}{P(y_{l+1}^k)}\h)^\frac{1}{1+\gamma}P(y_{l+1}^k) \\
        & = & \sum_{y_{l+1}^k} P(y_{l+1}^k) \sum_{x_{l+1}^k}
            R(x_{l+1}^k)\h(\frac{W(y_{l+1}^k|x_{l+1}^k)}{P(y_{l+1}^k)}\h)^\frac{1}{1+\gamma} \\
        & = & \h( \sum_{y\in \mY} P(y) \sum_{x \in \mX} \Beta(x) \B(\frac{W(y|x)}{P(y)}
            \B)^\frac{1}{1+\gamma} \h)^{k-l}
    \end{eqnarray}

So finally for $k \geq l$, using the definitions of $G_{si}$ and $G$ gives
     \begin{eqnarray}
        A_{l,k} & = & \exp \h( (l-k)\frac{\gamma}{1+\gamma} G + l E_{si}(\gamma) + (k-l) G_{si}(\gamma)
        - lE_0(\gamma) - (k-l)G(\gamma) \h)
     \end{eqnarray}

Now we split the double sum in the bound of $E[N^\gamma]$ and use the two cases of $l,k$ to
get:
    \begin{eqnarray}
        E[N^\gamma] & \leq & \sum_{l=1}^\infty \sum_{k=1}^\infty  A_{l,k}\\
        & = & \sum_{l=1}^\infty \sum_{k=l}^\infty A_{l,k} + \sum_{k=1}^\infty \sum_{l=k+1}^\infty
            A_{l,k}\\
        & \leq & \sum_{l=1}^\infty \sum_{k=l}^\infty A_{l,k} + \sum_{k=1}^\infty \sum_{l=k}^\infty
            A_{l,k} \\
        & \leq & \sum_{l=1}^\infty \sum_{k=l}^\infty \exp \h( (l-k)\frac{\gamma}{1+\gamma} G + l E_{si}(\gamma) + (k-l) G_{si}(\gamma)
            - lE_0(\gamma) - (k-l)G(\gamma) \h) + \nonumber \\
        & & \hspace{.7cm} \sum_{k=1}^\infty \sum_{l=k}^\infty \exp \B( (l-k)\frac{\gamma}{1+\gamma}G +
            k E_{si}(\gamma) + (l-k) F_{si}(\gamma) - k E_0(\gamma) - (l-k) F(\gamma)\h) \\
        & = & \sum_{l=1}^\infty \exp\b( l(E_{si}(\gamma) - E_0(\gamma))\b)\sum_{k=l}^\infty \exp \h(
            (l-k)\frac{\gamma}{1+\gamma} G + (k-l) G_{si}(\gamma)  - (k-l)G(\gamma) \h) \\
        & & \hspace{.2cm}+\sum_{k=1}^\infty \exp\b( k(E_{si}(\gamma) - E_0(\gamma))\b) \sum_{l=k}^\infty
            \exp \B( (l-k)\frac{\gamma}{1+\gamma}G + (l-k) F_{si}(\gamma)  - (l-k) F(\gamma)
            \B)\nonumber
    \end{eqnarray}

Now, if $\lambda \neq 1$, we would instead have
    \begin{eqnarray}
        E[N^\gamma] & \leq & \sum_{l=1}^\infty \exp\b( l(E_{si}(\gamma) - \lambda E_0(\gamma))\b)\sum_{k=l}^\infty \exp \h(
            (l-k)\frac{\gamma}{1+\gamma} G + (k-l) G_{si}(\gamma)  - \lambda (k-l)G(\gamma) \h) \\
        & & \hspace{.2cm}+\sum_{k=1}^\infty \exp\b( k(E_{si}(\gamma) - \lambda E_0(\gamma))\b) \sum_{l=k}^\infty
            \exp \B( (l-k)\frac{\gamma}{1+\gamma}G + (l-k) F_{si}(\gamma)  - \lambda (l-k) F(\gamma)
            \B)\nonumber
    \end{eqnarray}
 The above sums converge if the following
conditions are met:
    \begin{eqnarray}
        E_{si}(\gamma) & < & \lambda E_0(\gamma) \label{eqn:one}\\
        \frac{\gamma}{1+\gamma}G & > & G_{si}(\gamma) - \lambda G(\gamma) \label{eqn:two}\\
        \frac{\gamma}{1+\gamma}G & < & \lambda F(\gamma) - F_{si}(\gamma) \label{eqn:three}
    \end{eqnarray}

Condition (\ref{eqn:one}) is effectively the requirement that the source coding computational
cutoff rate for the $\gamma^{th}$ moment is lower than the channel coding cutoff rate for the
$\gamma^{th}$ moment. This is needed in this case {\em even though we are using joint
source-channel coding}. Conditions (\ref{eqn:two}) and (\ref{eqn:three}) combined require
    \begin{eqnarray}
        \frac{1+\gamma}{\gamma} \B[G_{si}(\gamma) - \lambda G(\gamma) \B] < & G & <\frac{1+\gamma}{\gamma}\B[
        \lambda F(\gamma) - F_{si}(\gamma) \B]
    \end{eqnarray}
\end{proof}

\subsection{Showing the range of viable bias values is non-empty} \label{sec:nonemptyrange}

Fix a $\gamma \in [0,1]$. For each $v\in \mV$ and $y\in \mY$ define $H(v)$ and $J(y)$ as:
    \begin{eqnarray}
        H(v) & \triangleq & \sum_{u \in \mU} Q(u|v)^\frac{1}{1+\gamma} \\
        J(y) & \triangleq & \sum_{x\in \mX} \Beta(x) \B( \frac{W(y|x)}{P(y)}\B)^\frac{1}{1+\gamma}
    \end{eqnarray}

If we consider $V$ to a random variable with distribution $Q(v)$ on $\mV$ and $Y$ to be a
random variable with distribution $P(y)$ on $\mY$, then by definition we have the following
relations:
    \begin{eqnarray}
        E_{si}(\gamma) & = & \log_2 E[H(V)^{1+\gamma}] \\
        F_{si}(\gamma) & = & \log_2 E[H(V)^\gamma]\\
        G_{si}(\gamma) & = & \log_2 E[H(V)]\\
        E_0(\gamma) & = & -\log_2 E[J(Y)^{1+\gamma}] \\
        F(\gamma) & = & -\log_2 E[J(Y)^\gamma]\\
        G(\gamma) & = & -\log_2 E[J(Y)]
    \end{eqnarray}

By repeated use of Jensen's inequality, since $\gamma \in [0,1]$, we also have
    \begin{eqnarray}
        E[H(V)^{1+\gamma}] & \geq & E[H(V)]E[H(V)]^\gamma \\
        & \geq & E[H(V)]E[H(V)^\gamma] \\
        E[J(Y)^{1+\gamma}] & \geq & E[J(Y)]E[J(Y)]^\gamma \\
        & \geq & E[J(Y)]E[J(Y)^\gamma]
    \end{eqnarray}

Since $\log_2$ is a monotonically increasing function, this means:
    \begin{eqnarray}
        E_{si}(\gamma) & \geq & F_{si}(\gamma) + G_{si}(\gamma) \\
        E_0(\gamma) & \leq & F(\gamma) + G(\gamma)
    \end{eqnarray}

Now, if $\gamma R > E_{si}(\gamma)$, then
    \begin{eqnarray}
        \gamma R - F_{si}(\gamma) - G_{si}(\gamma) & > & E_{si}(\gamma) - F_{si}(\gamma) -
        G_{si}(\gamma) \\
        & \geq & 0
    \end{eqnarray}
Hence, $(\frac{1+\gamma}{\gamma}G_{si}(\gamma), \frac{1+\gamma}{\gamma}(\gamma R -
F_{si}(\gamma)))$ is a non-empty open interval of bias values that give a finite $\gamma^{th}$
moment of computation if $\gamma R > E_{si}(\gamma)$ as shown in section \ref{sec:compsi}.

For the joint source-channel case, we assume $\lambda E_0(\gamma) > E_{si}(\gamma)$. Then,
    \begin{eqnarray}
        \lambda F(\gamma) + \lambda G(\gamma) - F_{si}(\gamma) - G_{si}(\gamma) & \geq & \lambda  E_0(\gamma) - E_{si}(\gamma) \\
        & > & 0
    \end{eqnarray}

Hence, there is a non-empty open interval of allowable bias values in Theorem \ref{thm:compsc}
if $\lambda E_0(\gamma) > E_{si}(\gamma)$.
\subsection{Error exponent with bias set for computation} \label{sec:biasforcomp}
    In this section, it is shown that if the bias can be set to achieve a $\gamma^{th}$ moment
of computation while still allowing for a positive error exponent.

    In the source coding with side information case, assume $\gamma R > E_{si}(\gamma)$, then we
know (Thm. \ref{thm:proberrorsi}) that for all $\eps> 0$, there is a $K_\eps < \infty$ so that
    \begin{equation}
        P_e(d) < K_\eps 2^{-d(\gamma R - E_{si} (\gamma) - \eps)}
    \end{equation}

This is provided that the bias $G$ satisfies
    \begin{equation}
        G \leq \frac{1+\gamma}{\gamma} [ E_{si}(\gamma) - F_{si}(\gamma) ]
    \end{equation}

Also, from Thm. \ref{thm:compsi}, the $\gamma^{th}$ moment of computation is finite provided
    \begin{equation}
        \frac{1+\gamma}{\gamma} G_{si}(\gamma) < G < \frac{1+\gamma}{\gamma} [ \gamma R -
        F_{si}(\gamma)]
    \end{equation}

Suppose the bias is set so that $G^* = \frac{1+\gamma}{\gamma} [ E_{si}(\gamma) -
F_{si}(\gamma)]$. Then there is a positive error exponent with delay. It is also true, however,
that this choice of bias yields a finite $\gamma^{th}$ moment of computation. Since we assume
$\gamma R > E_{si}(\gamma)$, it is immediate that $G^* < \frac{1+\gamma}{\gamma}[\gamma R -
F_{si}(\gamma)]$.

For the other inequality, we need that the $\log$ function is strictly concave $\cap$. This
combined with the assumption that $U$ is not deterministic given $v \in \mV$ for at least one
$\mV$ gives the strict inequality below:
    \begin{equation}
        G_{si}(\gamma) + F_{si}(\gamma) < E_{si}(\gamma)
    \end{equation}

Hence, $G^* > \frac{1+\gamma}{\gamma} G_{si}(\gamma)$ if the source $U$ is not deterministic
given $v$ for at least one value of $v \in \mV$ \footnote{If $U$ is conditionally deterministic
given $v$ for all $v \in \mV$, obviously the source coding with side information problem is not
interesting as zero rate is needed.}.

    For the joint source-channel coding with side information case, an analogous line of reasoning
gives that the choice $G^* = \frac{1+\gamma}{\gamma} [E_{si}(\gamma) - F_{si}(\gamma) - \lambda
E_0(\gamma) + \lambda F(\gamma)]$ gives a positive error exponent and finite $\gamma^{th}$
moment of computation provided $E_{si}(\gamma) < E_0(\gamma)$.

\end{document}